\documentclass[11pt,letterpaper]{article}

\IfFileExists{latexml.sty}{\usepackage{latexml}}{\newif\iflatexml\latexmlfalse}

\usepackage[T1]{fontenc}
\usepackage[utf8]{inputenc}
\iflatexml
\newcommand{\carlito}{}\newcommand{\caladea}{}
\else
\usepackage{lmodern}                 %
\usepackage{caladea}                 %
\usepackage[lining]{carlito}         %
\usepackage[default,lining]{lato}    %
\DeclareFontFamily{T1}{LatoW}{}
\DeclareFontShape{T1}{LatoW}{m}{n}{<-> Lato-Regular-T1-TLF--base}{}
\DeclareFontShape{T1}{LatoW}{m}{it}{<-> Lato-Italic-T1-TLF--base}{}
\DeclareFontShape{T1}{LatoW}{b}{n}{<-> Lato-Bold-T1-TLF--base}{}
\DeclareFontShape{T1}{LatoW}{b}{it}{<-> Lato-BoldItalic-T1-TLF--base}{}
\DeclareFontShape{T1}{LatoW}{m}{sl}{<-> ssub * LatoW/m/it}{}
\DeclareFontShape{T1}{LatoW}{b}{sl}{<-> ssub * LatoW/b/it}{}
\DeclareFontShape{T1}{LatoW}{bx}{n}{<-> ssub * LatoW/b/n}{}
\DeclareFontShape{T1}{LatoW}{bx}{it}{<-> ssub * LatoW/b/it}{}
\DeclareFontShape{T1}{LatoW}{bx}{sl}{<-> ssub * LatoW/b/it}{}
\DeclareFontFamily{TS1}{LatoW}{}
\DeclareFontShape{TS1}{LatoW}{m}{n}{<-> ssub * lato-TLF/m/n}{}
\DeclareFontShape{TS1}{LatoW}{m}{it}{<-> ssub * lato-TLF/m/it}{}
\DeclareFontShape{TS1}{LatoW}{b}{n}{<-> ssub * lato-TLF/b/n}{}
\DeclareFontShape{TS1}{LatoW}{b}{it}{<-> ssub * lato-TLF/b/it}{}
\DeclareFontShape{TS1}{LatoW}{bx}{n}{<-> ssub * lato-TLF/b/n}{}
\DeclareFontShape{TS1}{LatoW}{bx}{it}{<-> ssub * lato-TLF/b/it}{}

\DeclareFontFamily{T1}{CrltW}{}
\DeclareFontShape{T1}{CrltW}{m}{n}{<-> Crlt-tlf-t1--base}{}
\DeclareFontShape{T1}{CrltW}{m}{it}{<-> Crlt-Italic-tlf-t1--base}{}
\DeclareFontShape{T1}{CrltW}{b}{n}{<-> Crlt-Bold-tlf-t1--base}{}
\DeclareFontShape{T1}{CrltW}{b}{it}{<-> Crlt-BoldItalic-tlf-t1--base}{}
\DeclareFontShape{T1}{CrltW}{bx}{n}{<-> ssub * CrltW/b/n}{}
\DeclareFontShape{T1}{CrltW}{bx}{it}{<-> ssub * CrltW/b/it}{}
\renewcommand*{\carlito}{\fontfamily{CrltW}\selectfont}
\usepackage{textcomp}
\usepackage[protrusion=false,expansion=false]{microtype} %
\DisableLigatures[f]{encoding = *, family = *}  %
\input{glyphtounicode}\pdfgentounicode=1
\pdfglyphtounicode{fi}{0066 0069}\pdfglyphtounicode{f_i}{0066 0069}\pdfglyphtounicode{fl}{0066 006C}\pdfglyphtounicode{f_l}{0066 006C}
\pdfglyphtounicode{ff}{0066 0066}\pdfglyphtounicode{f_f}{0066 0066}\pdfglyphtounicode{ffi}{0066 0066 0069}\pdfglyphtounicode{f_f_i}{0066 0066 0069}
\pdfglyphtounicode{uniFB00}{0066 0066}\pdfglyphtounicode{uniFB01}{0066 0069}\pdfglyphtounicode{uniFB02}{0066 006C}\pdfglyphtounicode{uniFB03}{0066 0066 0069}\pdfglyphtounicode{uniFB04}{0066 0066 006C}
\pdfglyphtounicode{f_t.liga}{0066 0074}\pdfglyphtounicode{f_f_t.liga}{0066 0066 0074}\pdfglyphtounicode{f_dotlessi.liga}{0066 0131}
\fi

\usepackage[letterpaper,left=56.3bp,right=56.3bp,top=51.6bp,bottom=51.2bp,
            headheight=14bp,headsep=18bp,footskip=23.2bp]{geometry}
\usepackage[table]{xcolor}
\usepackage{graphicx}
\usepackage{array}
\usepackage{longtable}
\usepackage{enumitem}
\usepackage{framed}
\usepackage{titlesec}
\iflatexml\else\usepackage[titles]{tocloft}\fi
\usepackage{fancyhdr}
\usepackage{lastpage}
\usepackage{caption}
\usepackage{float}
\iflatexml\else\usepackage{needspace}\fi
\iflatexml\else\usepackage[absolute]{textpos}\fi
\usepackage[document]{ragged2e}
\usepackage[nocompress,nosort]{cite}  %
\usepackage{url}
\usepackage{etoolbox}
\AtBeginEnvironment{itemize}{\listsize}
\usepackage[hidelinks]{hyperref}
\hypersetup{
  pdftitle={Provable Authority: A Protocol for Verifiable Authority Resolution and Control Across Human, Organizational, and Agentic Transactions},
  pdfauthor={Jeremy Blackburn, Tim Hines, Matthew Koepp, Russell Lessard, Jason Patrick},
  pdfsubject={ChainIT Authority Protocol architecture and source basis (CIT-PA-WP-7.0)},
  pdfkeywords={agentic commerce, AI agents, delegated authority, payment authorization, verifiable credentials, transaction binding, zero trust, stablecoin, ChainIT}
}

\definecolor{body}{HTML}{1F2937}
\definecolor{cnavy}{HTML}{1B2A4A}
\definecolor{cblue}{HTML}{2B6CB0}
\definecolor{cgray}{HTML}{4A5568}
\definecolor{hgray}{HTML}{6B7280}
\definecolor{cteal}{HTML}{0F7C7B}
\definecolor{crule}{HTML}{D3DCE3}
\definecolor{thnavy}{HTML}{1B2A4A}\definecolor{trule}{HTML}{D4DCE4}
\definecolor{rowalt}{HTML}{F5F7FA}
\definecolor{gbar}{HTML}{2D7D46}\definecolor{gtitle}{HTML}{2D7D46}\definecolor{gfill}{HTML}{EAF4ED}
\definecolor{rbar}{HTML}{9B2C2C}\definecolor{rfill}{HTML}{FBECEC}
\definecolor{bbar}{HTML}{2B6CB0}\definecolor{bfill}{HTML}{EAF2F8}
\definecolor{ybar}{HTML}{B8891E}\definecolor{yfill}{HTML}{FBF3DF}\definecolor{ytext}{HTML}{B8891E}
\definecolor{ffill}{HTML}{F5F7F9}

\renewcommand{\normalsize}{\fontsize{10.5bp}{14.23bp}\selectfont}
\newcommand{\boxsize}{\fontsize{9.5bp}{12.8bp}\selectfont}
\newcommand{\listsize}{\fontsize{10bp}{13.65bp}\selectfont}

\newcommand{\refsize}{\fontsize{8bp}{10.4bp}\selectfont}
\AtBeginDocument{\normalsize\color{body}}
\iflatexml\else\setlist[itemize]{label=\textbullet,leftmargin=17.2bp,labelindent=6.5bp,labelsep=*,itemsep=0bp,parsep=0bp,topsep=0.3bp,partopsep=0bp,after=\vspace{-3.1bp},before={\clubpenalty=0\widowpenalty=0\csname @clubpenalty\endcsname=0 \csname @itempenalty\endcsname=0 }}\fi %
\newif\ifpanellast\newif\iftablelast
\iflatexml\else\AddToHook{para/begin}{\global\panellastfalse\global\tablelastfalse}\fi

\titleformat{\section}{\raggedright\hyphenpenalty=10000\exhyphenpenalty=10000\carlito\bfseries\fontsize{16bp}{22.3bp}\selectfont\color{cblue}}{\rule{0bp}{15.4bp}\thesection.}{0.25em}{\rule{0bp}{15.4bp}}
\titleformat{\subsection}{\raggedright\hyphenpenalty=10000\carlito\bfseries\fontsize{13bp}{16bp}\selectfont\color{cblue}}{\rule{0bp}{12.7bp}\thesubsection}{0.25em}{\rule{0bp}{12.7bp}}
\titlespacing*{\section}{0bp}{\ifpanellast 15bp\else\iftablelast 14.1bp\else 4.5bp\fi\fi}{2.1bp}
\titlespacing*{\subsection}{0bp}{\ifpanellast 15.1bp\else\iftablelast 14.6bp\else 5bp\fi\fi}{1bp}
\newcommand{\tealhead}[1]{\par\vspace{\iftablelast 6.1bp\else 2bp\fi}{\carlito\bfseries\fontsize{11bp}{13.5bp}\selectfont\color{cteal}#1}\par\nobreak\vspace{-1.8bp}}
\newcommand{\runin}[1]{\textbf{#1}}
\newcommand{\unnumbered}[1]{\section*{#1}\addcontentsline{toc}{section}{#1}}

\iflatexml\else

\renewcommand{\cftsecpagefont}{}
\renewcommand{\cftdotsep}{0}%
\renewcommand{\cftsecfillnum}[1]{{\cftsecleader}\nobreak\hbox{\kern1bp\cftsecpagefont #1}\cftsecafterpnum\par}
\renewcommand{\cftsubsecfillnum}[1]{{\cftsubsecleader}\nobreak\hbox{\kern1bp\cftsubsecpagefont #1}\cftsubsecafterpnum\par}
\renewcommand{\cftsecleader}{\cftdotfill{\cftdotsep}}
\newsavebox{\tocnum}
\renewcommand{\cftsecpresnum}{\global\setbox\tocnum\hbox\bgroup}
\renewcommand{\cftsecaftersnum}{.\egroup}

\renewcommand{\cftsubsecpresnum}{\global\setbox\tocnum\hbox\bgroup}
\renewcommand{\cftsubsecaftersnum}{\egroup}

\fi

\DeclareCaptionFont{navy}{\color{cnavy}}
\DeclareCaptionFont{capsize}{\fontsize{8.5bp}{10.5bp}\selectfont}
\newcommand{\fignote}[1]{\par\vspace{-2bp}{\fontsize{8bp}{9.6bp}\selectfont\itshape\color{cgray}#1\par}}

\iflatexml
\newenvironment{callout}[5]{\begin{quote}\ifx&#4&\else\textbf{#4}\par\fi}{\end{quote}}
\newenvironment{greenbox}[1]{\begin{callout}{}{}{}{#1}{}}{\end{callout}}
\newenvironment{redbox}[1]{\begin{callout}{}{}{}{#1}{}}{\end{callout}}
\newenvironment{bluebox}[1]{\begin{callout}{}{}{}{#1}{}}{\end{callout}}
\newcommand{\goldlead}[2][]{\begin{quote}\textbf{#2}\end{quote}}
\newenvironment{formulabox}[1]{\begin{quote}\centering\textbf{#1}\par}{\end{quote}}
\newcommand{\formula}[1]{\textbf{#1}\par}
\newcommand{\formulanote}[1]{\emph{#1}\par}
\else
\newsavebox{\fbbox}
\newenvironment{callout}[5]%
  {\par\ifdim\lastskip=0bp\vspace{-2.8bp}\fi%
   \setlist*[itemize]{topsep=0bp,partopsep=0bp,after=}%
   \vbadness=10000 %
   \def\FrameCommand##1{\sbox{\fbbox}{\fboxsep=0bp\colorbox{#2}{\hspace*{7.3bp}\vbox{\kern4.5bp\hbox{##1}\kern7.3bp}\hspace*{7.3bp}}}%
     \hspace*{15bp}\raisebox{\dimexpr5.3bp-\dp\fbbox\relax}{\color{#1}\rule{#5}{\dimexpr\ht\fbbox+\dp\fbbox-10.6bp\relax}}%
     \usebox{\fbbox}\hspace*{14.7bp}}%
   \MakeFramed{\advance\hsize-\width\FrameRestore}%
   \boxsize\setlength{\parskip}{3bp}\clubpenalty=0\widowpenalty=0%
   \ifx&#4&\else{\bfseries\fontsize{10bp}{12.5bp}\selectfont\color{#3}#4}\par\vspace{-0.5bp}\fi}
  {\endMakeFramed\global\panellasttrue}
\newenvironment{greenbox}[1]{\begin{callout}{gbar}{gfill}{gtitle}{#1}{6.7bp}}{\end{callout}}
\newenvironment{redbox}[1]{\begin{callout}{rbar}{rfill}{rbar}{#1}{7.2bp}}{\end{callout}}
\newenvironment{bluebox}[1]{\begin{callout}{bbar}{bfill}{cblue}{#1}{7.2bp}}{\end{callout}}
\newcommand{\goldlead}[2][0bp]{\par\vspace{-2.8bp}%
  {\def\FrameCommand##1{\sbox{\fbbox}{\fboxsep=0bp\colorbox{yfill}{\hspace*{5bp}\vbox{\kern-0.5bp\hbox{##1}\kern\dimexpr5bp+#1\relax}\hspace*{5bp}}}%
     \raisebox{\dimexpr3bp-\dp\fbbox\relax}{\color{ybar}\rule{6.5bp}{\dimexpr\ht\fbbox+\dp\fbbox-6bp\relax}}%
     \usebox{\fbbox}\hspace*{26.2bp}}%
   \MakeFramed{\advance\hsize-\width\FrameRestore}%
   {\bfseries\fontsize{8bp}{10bp}\selectfont\color{ytext}#2\par}\endMakeFramed}}

\newenvironment{formulabox}[1]
  {\par\vspace{-6.8bp}%
   \def\FrameCommand{\hspace*{17.2bp}\fboxsep=7bp\fboxrule=1bp\fcolorbox{crule}{ffill}}%
   \MakeFramed{\advance\hsize-\width\advance\hsize-16.7bp\FrameRestore}%
   \centering\setlength{\parskip}{5bp}%
   {\bfseries\fontsize{9bp}{11bp}\selectfont\color{cblue}#1}\par}
  {\endMakeFramed\vspace{1.9bp}}
\newcommand{\formula}[1]{{\caladea\bfseries\boldmath\fontsize{9bp}{12bp}\selectfont\color{cnavy}#1\par}}
\newcommand{\formulanote}[1]{{\fontsize{8.5bp}{11.5bp}\selectfont\itshape\color{cgray}#1\par}}
\fi

\arrayrulecolor{trule}
\newcommand{\tabsize}{\fontsize{8bp}{10.85bp}\selectfont}
\newlength{\padtop}\newlength{\padbot}
\newcommand{\tablestyle}[4]{\renewcommand{\tabsize}{\fontsize{#1}{#2}\selectfont}\setlength{\padtop}{#3}\setlength{\padbot}{#4}}
\newcommand{\cellpadtop}{\rule{0bp}{\padtop}}
\newcommand{\cellpadbot}{\rule[-\padbot]{0bp}{0bp}}
\iflatexml\else\AfterEndEnvironment{longtable}{\global\tablelasttrue}\fi
\newcolumntype{M}[1]{>{\RaggedRight\arraybackslash\tabsize\cellpadtop}m{#1}<{\cellpadbot}}
\newcolumntype{B}[1]{>{\RaggedRight\arraybackslash\tabsize\bfseries\color{cnavy}\cellpadtop}m{#1}<{\cellpadbot}}
\newcolumntype{D}[1]{>{\RaggedRight\arraybackslash\tabsize\cellpadtop}m{#1}<{\cellpadbot}}
\newcolumntype{Y}[1]{>{\RaggedRight\arraybackslash\fontsize{8.5bp}{11.5bp}\selectfont\cellpadtop}m{#1}<{\cellpadbot}}
\newcommand{\thc}[1]{\rule[-\dimexpr\padbot+0.3bp\relax]{0bp}{\dimexpr\padtop+0.4bp\relax}\textcolor{white}{\bfseries\tabsize #1}}
\newcommand{\headrow}{\rowcolor{thnavy}}
\newcommand{\altrow}{\rowcolor{rowalt}}

\title{Provable Authority: A Protocol for Verifiable Authority Resolution and Control Across Human, Organizational, and Agentic Transactions}
\newcommand{\affil}{ChainIT, Utah, United States (\url{https://chainit.com})}%
\author{Jeremy Blackburn\thanks{\affil} \and Tim Hines\thanks{\affil} \and Matthew Koepp\thanks{\affil} \and Russell Lessard\thanks{\affil} \and Jason Patrick\thanks{\affil}}
\date{September 1, 2026}

\iflatexml\else
\renewenvironment{abstract}{\section*{Abstract}}{\par}
\fi

\begin{document}

\iflatexml
\maketitle
\else
\thispagestyle{empty}
\setlength{\TPHorizModule}{1bp}\setlength{\TPVertModule}{1bp}
\newcommand{\coverline}[3]{%
  \begin{textblock*}{499.4bp}(56.3bp,#1bp)\centering #2 #3\par\end{textblock*}}
\coverline{61.2}{\bfseries\fontsize{30bp}{36bp}\selectfont\color{cnavy}}{Provable Authority}
\coverline{104.3}{\fontsize{12.5bp}{17bp}\selectfont\color{cgray}}{A Protocol for Verifiable Authority Resolution and Control Across Human, Organizational,\\and Agentic Transactions}
\coverline{214.5}{\bfseries\fontsize{15bp}{18bp}\selectfont\color{cgray}}{ChainIT}
\coverline{239.8}{\bfseries\fontsize{12bp}{14.5bp}\selectfont\color{cblue}}{\url{https://chainit.com}}
\coverline{331.1}{\bfseries\fontsize{11bp}{13.5bp}\selectfont\color{cnavy}}{ChainIT Authority Protocol}
\coverline{351.5}{\itshape\fontsize{9.5bp}{11.5bp}\selectfont\color{cgray}}{Architecture and Source Basis}
\coverline{474.3}{\fontsize{10.5bp}{12.5bp}\selectfont\color{hgray}}{September 1, 2026}
\coverline{525.8}{\bfseries\fontsize{10.5bp}{12.5bp}\selectfont\color{body}}{Authors}
\coverline{545.8}{\fontsize{9.5bp}{13bp}\selectfont\color{body}}{\href{https://chainit.com/about-us/}{Jeremy Blackburn}, Chief Executive Officer\\\href{https://chainit.com/about-us/}{Tim Hines}, Chief Product Officer\\\href{https://chainit.com/about-us/}{Matthew Koepp}, Chief Technology Officer\\\href{https://chainit.com/about-us/}{Russell Lessard}, Chief Compliance Officer\\\href{https://chainit.com/about-us/}{Jason Patrick}, Chief Commercial Officer}
\coverline{641.2}{\fontsize{8bp}{10.7bp}\selectfont\color{cgray}}{This paper describes the ChainIT \href{https://chainit.com/insights/chainit-launches-provable-authority-as-u-s-generative-ai-enabled-fraud-losses-are-projected-to-reach-40-billion-by-2027/}{Authority Protocol} architecture. It does not constitute legal, regulatory, banking, accounting, security-\\certification, or patentability advice. Deployment configurations, partner rail support, and service levels vary by implementation.}
\coverline{677.2}{\bfseries\fontsize{11bp}{13bp}\selectfont\color{black}}{\textcopyright{} 2026, Scientia Potentia Est.\ II, LLC. All rights reserved.}
\coverline{694.6}{\fontsize{8bp}{9.6bp}\selectfont\color{cgray}}{ChainIT is a registered trademark of Scientia Potentia Est.\ II, LLC. Distributed by arXiv under the arXiv.org perpetual, non-exclusive license.}
\null\clearpage
\fi
\color{body}

\begin{abstract}
Payment systems authenticate users, devices, accounts, and cryptographic credentials, but a relying party, such as a counterparty, payment rail, marketplace, or autonomous agent runtime, generally cannot obtain compact, independently verifiable evidence that the party presenting one exact transaction holds current organizational authority to execute it under the applicable evidence, policy, budget, and approval state. Agentic commerce makes this gap operationally urgent: a software agent can produce a valid signature while acting outside its delegated purpose, after its sponsor's authority has changed, or on transaction details that differ from those a human approved. We describe the ChainIT \href{https://chainit.com/insights/chainit-launches-provable-authority-as-u-s-generative-ai-enabled-fraud-losses-are-projected-to-reach-40-billion-by-2027/}{Authority Protocol}, a subject-neutral and rail-neutral pre-execution authority and evidence control plane. The protocol separates durable identity, a versioned organizational authority record (the Authority Resolution \href{https://www.pactvera.com/}{Pactvera}), sourced evidence assertions, deterministic reason-coded policy predicates, and organization-configured dispositions; binds each decision and human approval to a canonical transaction digest; reserves capacity atomically so that limits hold under concurrency; and authorizes commit only through a single-use, sender-constrained Execution Authorization Credential consumed at the rail adapter, with append-only lifecycle evidence sealed after the terminal outcome. An Agent Subject Profile adds the workload registration, attested key hierarchy, and sovereignty boundary required when the acting subject is an AI agent, without introducing a separate authority concept. We specify the commit-time execution predicate and bounded-freshness status rules, give enforcement profiles for fiat (ACH, wire, card, instant payment) and stablecoin rails, and state residual risks and non-goals.
\end{abstract}


{\fontsize{9.5bp}{12.8bp}\selectfont\color{cgray}This paper was first released as a ChainIT white paper (CIT-PA-WP-7.0), available at \url{https://chainit.com/chainit_provable_authority_whitepaper/}. The companion paper, \emph{Provable Compliance} \cite{provable-compliance}, and other ChainIT white papers are listed at \url{https://chainit.com/white-papers/}.\par}
\color{body}
\clearpage

\iflatexml\else
\addcontentsline{toc}{section}{Contents}
\tableofcontents
\clearpage
\fi

\unnumbered{Definitions}\label{sec:definitions}
\vspace{6bp}%

{\color{black}Terms below are used as defined here throughout this paper. The \href{https://chainit.com/insights/chainit-launches-provable-authority-as-u-s-generative-ai-enabled-fraud-losses-are-projected-to-reach-40-billion-by-2027/}{Authority Protocol} is the general authority and pre-execution control protocol and applies to every subject of authority and to fiat, card, instant-payment, stablecoin, and on-chain transactions alike. The Agent Subject Profile is the narrower profile covering the identity and containment controls that exist only because the acting subject is an autonomous workload. Definitions are provided for reader clarity and do not themselves constitute legal, regulatory, or product specifications.\par}%

{\setlength{\parskip}{11.5bp}

\runin{Agent Authority \href{https://chainit.com/product/data-tokenization/}{VDT}.} The append-only record of a grant of delegated authority from an authorized human or organizational principal to a registered agent, and the agent-specialized form of the Delegated Authority VDT. Each entry immutably records the granting principal, the governing ARP and its version, the registered agent, the permitted purpose, counterparties, transaction classes, rails, per-transaction and aggregate value limits, effective period, step-up thresholds, redelegation rule, and revocation conditions. An agent holds no authority of its own; the Agent Authority VDT is the provable evidence that authority was granted, by whom, and within what bounds. The record is immutable, but present validity is resolved at the reliance event against current ARP status and the applicable freshness bound. It is not a bearer credential.

\runin{Agent Registration \href{https://chainit.com/product/data-tokenization/}{VDT}.} A versioned record identifying an autonomous workload\textquotesingle{}s organization, responsible party or role, tenant, deployment, software or workload identity, permitted tools and APIs, risk class, attestation method, registered public key, status reference, and version history. It establishes what the workload is, not what it may do.

\runin{Agent Subject Profile.} The identity and containment controls that exist because the acting subject is an autonomous workload rather than a person: the Agent Registration \href{https://chainit.com/product/data-tokenization/}{VDT}, the attested workload key and ephemeral session key hierarchy, and the sovereignty boundary. The profile adds no authority concept of its own (an agent exercises delegated authority under the ARP in the same way a human delegate does, evidenced by an Agent Authority VDT).

\runin{Authority Policy Profile.} The versioned rule set selected for a given reliance event, which may vary by organization, product, role, subject class, counterparty, transaction type, value, payment rail, jurisdiction, risk tier, and intended purpose. It is the authority-side counterpart to the Compliance Policy Profile described in ChainIT \href{https://chainit.com/white-papers/}{Provable Compliance}, and it is versioned so that a later reviewer can determine exactly which rule set produced a decision.

\runin{Authority Protocol.} The general ChainIT protocol for resolving whether an identified subject may perform a specific proposed action at the moment of reliance. It is not limited to agentic commerce and is not rail-specific: the same identity, authority, evidence, policy, digest, reservation, execution, and outcome-evidence controls govern a wire a finance director releases, a dual-approved ACH file, a card or instant-payment instruction, a stablecoin transfer, and an agent-initiated proposal.

\runin{Authority Resolution \href{https://www.pactvera.com/}{Pactvera} (ARP).} The canonical, versioned organizational authority record: who may act, in what role, for what purpose, within what limits, during what effective period, under what quorum or separation-of-duties requirements, and subject to what amendment or revocation.

\runin{Business Rules Engine (BRE).} The deterministic predicate-evaluation engine. It evaluates configured conditions against subject identity and assurance, ARP authority, registration and attestation where the subject is a workload, evidence, status freshness, jurisdiction, scope, session risk, and for a transaction, the exact transaction, current capacity, required approvals, and rail-specific constraints, and returns explicit, reason-coded true/false predicate results. The BRE does not itself perform risk scoring or assign a disposition; it runs within the ChainIT Workflow Engine, where the organization configures how predicate results combine into a disposition.

\runin{Canonical transaction digest.} A deterministic hash of a canonical transaction object (payer, source, payee, destination, amount, currency or asset, rail, fees, nonce, validity window, and rail-specific parameters, among other fields) computed under a fixed encoding rule so that any material change to the transaction produces a different digest.

\runin{ChainIT ID.} The verified identity anchor for an individual, established through authoritative-source verification, biometric and liveness controls, and privacy-preserving, portable identity storage.

\runin{ChainIT Org ID.} The legal-entity identity anchor for an organization, which may reference entity registration, ownership, officers, governance, and authorized representatives.

\runin{ChainIT Pay.} The ChainIT payment-execution platform intended to serve as the EAC enforcement boundary where the applicable \href{https://chainit.com/insights/chainit-launches-provable-authority-as-u-s-generative-ai-enabled-fraud-losses-are-projected-to-reach-40-billion-by-2027/}{Authority Protocol} profile is deployed; a compatible partner rail adapter may serve this function in another implementation.

\runin{Compliance Decision \href{https://chainit.com/product/data-tokenization/}{VDT}.} The signed, time-bounded, purpose-specific record produced by an organization's configuration of the ChainIT Workflow Engine, recording whether present evidence and policy permit a proposed exercise of authority. Defined in ChainIT \href{https://chainit.com/white-papers/}{Provable Compliance} and consumed here as the Compliance Decision Permits predicate. It gates present exercise; it does not amend, suspend, or erase the governing ARP.

\runin{Delegated Authority \href{https://chainit.com/product/data-tokenization/}{VDT}.} A portable artifact that references the governing ARP and exposes only the scope of authority needed by a given delegate, agent, or relying party, without replacing or superseding the ARP itself. Its agent-specialized form is the Agent Authority VDT.

\runin{Disposition.} The explicit outcome produced by an organization's configuration of the ChainIT Workflow Engine from BRE predicate results (e.g., Allow, Allow with Controls, Step-Up, Review, Hold, Reject, Prohibit, or Stale/Unknown). The disposition set is configured by the organization to match its own policies and regulatory obligations; it is not a risk judgment ChainIT renders on the organization's behalf. Disposition and risk tier are evaluated independently (a high-risk transaction may still resolve to Allow, and a low-risk transaction may still be rejected).

\runin{Evidence Assertion \href{https://chainit.com/product/data-tokenization/}{VDT}.} A record of a single sourced fact, such as identity-document status, sanctions-screening result, or entity status, including its subject, source, source version, retrieval and effective time, expiration, jurisdiction, match or confidence state, and privacy classification. Evidence Assertion VDTs are never aggregated into a single \textquotedbl{}compliant\textquotedbl{} status. This is the same artifact class as the Evidence VDT defined in ChainIT \href{https://chainit.com/white-papers/}{Provable Compliance}, which classifies each item as authoritative-source, qualified third-party, or institution-generated evidence and records what the source does and does not prove.

\runin{Execution Authorization Credential (EAC).} A single-use, short-lived credential authorizing one exact transaction for one audience, bound to the acting session or approved subject, transaction digest, reservation, status epoch, and expiration, and consumed atomically at commit.

\runin{Execution Authorization Lifecycle \href{https://chainit.com/product/data-tokenization/}{VDT}.} A permanent, append-only evidence record of EAC issuance and terminal handling (consumed, expired, canceled, or rejected). Earlier lifecycle events are never modified or overwritten.

\runin{Pactvera.} The ChainIT transaction-logic and append-only evidence framework linking \href{https://chainit.com/product/data-tokenization/}{Validated Data Tokens} across a transaction\textquotesingle{}s lifecycle.

\runin{Reliance event.} The moment at which a relying party depends on identity, authority, evidence, or policy state — session issuance or renewal, a transaction proposal or payment initiation, a human step-up approval, a scheduled review, or a source-change or material risk event. Freshness and policy are evaluated at the reliance event rather than inherited from onboarding.

\runin{Reservation.} The atomic hold of transaction capacity against a budget or limit, performed before Execution Authorization Credential issuance so that concurrent requests cannot commit against the same capacity twice.

\runin{Risk tier.} A classification (Tier 1–Low, Tier 2–Standard, Tier 3–High) assigned to a transaction based on factors such as payee history, value, reversibility, and behavioral or device anomalies, which determines the minimum acceptable verification rigor (e.g., cached status snapshot versus live status introspection). The tier is assigned by the organization\textquotesingle{}s configuration of the ChainIT Workflow Engine, not by the BRE; the assignment records the policy version applied, the input factors considered, the assignment time, applicable reason codes, and any manual override.

\runin{Session Authorization Credential (SAC).} A short-lived, sender-constrained credential that permits proposal activity, evidence retrieval, and defined API activity, but which rail adapters must reject as payment authority. A session credential is never payment authority, whether the session belongs to a workload or to a signed-in person.

\runin{Session Eligibility Decision \href{https://chainit.com/product/data-tokenization/}{VDT}.} A session-context decision recording whether the authenticated subject may open or continue a session that enables consequential action, referencing identity assurance, the ARP, evidence, status, jurisdiction, scope, and policy version. Where the subject is a workload, it additionally references the Agent Registration VDT, Agent Authority VDT, attested deployment, and applicable tool scope. It is not transaction-specific and cannot authorize payment.

\runin{Sovereignty boundary.} The architectural line beyond which an AI model or autonomous agent may reason, gather evidence, and propose an action, but cannot obtain unrestricted access to payment-account credentials, custody keys, or final execution authority, with that authority remaining with the deterministic control plane.

\runin{Status epoch / status freshness.} The signed timestamp or snapshot boundary used to determine whether cached non-revocation status is still acceptable for a given risk tier, versus requiring live status introspection.

\runin{Subject of authority.} The party whose authority is being resolved: a verified individual, a group acting under a quorum or separation-of-duties rule, an organization acting through an authorized representative, or a permitted software agent. The ARP, evidence, policy, and decision artifacts are the same across subject classes; what differs is how the subject is authenticated and registered, and what containment it requires. A scheduled workflow is not a fifth subject class; it is an organization-controlled initiation mechanism that acts on behalf of an identified organization and under the governing ARP, in the same manner as any other organizational or multi-party initiation.

\runin{Transaction Policy Decision \href{https://chainit.com/product/data-tokenization/}{VDT}.} A decision bound to one canonical transaction digest, recording risk tier, disposition, applicable policy version, evidence references, ARP and subject references (including agent references where the subject is a workload), required approvals or controls, expiration, and reason codes. It is newly derived per transaction and is not reusable as a standing status.

\runin{Validated Data Token (\href{https://chainit.com/product/data-tokenization/}{VDT}).} A cryptographically verifiable digital record representing validated data concerning a person, organization, asset, service, event, decision, or transaction. Within the \href{https://chainit.com/insights/chainit-launches-provable-authority-as-u-s-generative-ai-enabled-fraud-losses-are-projected-to-reach-40-billion-by-2027/}{Authority Protocol}, evidence and decision VDTs are immutable or versioned; a change is represented by a new linked VDT rather than alteration of a prior immutable record (e.g., Evidence Assertion VDT, Agent Registration VDT, Transaction Policy Decision VDT).

\runin{Valitorum.} The final sealed evidence record generated for a completed or otherwise terminal transaction lifecycle under a defined completeness rule. A sealed \href{https://chainit.com/}{Valitorum} is never altered; subsequent exceptions (returns, reversals, disputes) are captured in a linked Supplemental Valitorum that references the original.

\runin{Valitorum Service.} The ChainIT service that creates, verifies, retrieves, or presents a \href{https://chainit.com/}{Valitorum}.

\runin{Workflow Engine.} The ChainIT platform capability an organization configures to encode its own internal policies and regulatory requirements, orchestrating BRE predicate results into the disposition states and downstream actions the organization defines. The organization is responsible for the policy logic it configures and for its use of the resulting disposition, subject to applicable law and contractual allocation among ChainIT, the organization, payment providers, evidence providers, and other participating parties.

\runin{Workload key / session key.} The key hierarchy underlying agent operations: a rotatable workload key, generated or protected in an approved security boundary (e.g., TPM, TEE, KMS, HSM), authorizes an ephemeral session key; Session Authorization Credentials and Execution Authorization Credentials are sender-constrained to the session key rather than the workload key.

}
\clearpage

\unnumbered{Executive Summary}\label{sec:summary}

Payment systems authenticate users, devices, accounts, and cryptographic credentials. Financial institutions also enforce internal entitlements, limits, dual approval, sanctions screening, fraud controls, and account permissions. The persistent gap is portability: a counterparty, marketplace, payment rail, or autonomous workload generally cannot obtain compact, independently verifiable evidence that the party presenting one exact transaction possesses current organizational authority to do so under the applicable evidence, policy, budget, and approval state.

That gap is not created by autonomous agents and it is not limited to them. A finance director can hold a delegation that has since been amended. A dual-approval control can be satisfied on a screen that does not match the ACH file eventually transmitted. A wire can be released to a beneficiary account that changed after the invoice was approved. A per-transaction limit can be defeated by two concurrent requests reading the same balance. In each case the credential is authentic and the control is nominally present, but no relying party can independently verify that the exact instruction being committed is the one that current authority, evidence, and policy actually permit.

Agentic commerce makes that gap operationally urgent rather than novel. A software agent can produce a valid signature while acting outside its delegated purpose, after its human sponsor or organizational authority has changed, against stale or ambiguous evidence, beyond an aggregate budget, or on transaction details that differ from those a human approved. Cryptographic authenticity is necessary, but it does not answer the governance question: who stands behind the action, what authority was granted, whether that authority may be exercised now, and what precise instruction is being committed.

ChainIT \href{https://chainit.com/insights/chainit-launches-provable-authority-as-u-s-generative-ai-enabled-fraud-losses-are-projected-to-reach-40-billion-by-2027/}{Provable Authority} is a pre-execution authority and evidence control plane, grounded in the ChainIT platform, patent, and application materials cited throughout. \cite{chainit-identity}-[\citen{pat940}][\citen{chainitpay}] It does not replace payment rails. \href{https://chainit.com/product/identity-verification/}{ChainIT ID} and \href{https://chainit.com/product/business-verification/}{ChainIT Org ID} anchor verified subjects; a versioned Authority Resolution \href{https://www.pactvera.com/}{Pactvera} is the governing organizational authority record; the Business Rules Engine returns deterministic predicate results against a versioned Authority Policy Profile and sourced evidence; a transaction-specific decision binds an exact canonical transaction digest; capacity is reserved atomically; a one-time Execution Authorization Credential is consumed at the payment boundary; and Pactvera, append-only lifecycle \href{https://chainit.com/product/data-tokenization/}{VDTs} and \href{https://chainit.com/}{Valitorum} preserve what was decided, who acted, and what occurred. \href{https://chainit.com/product/digital-pay/}{ChainIT Pay}, or a compatible partner adapter, serves as the execution boundary where the applicable enforcement profile is deployed. These controls are subject-neutral and rail-neutral: they apply equally to a wire a finance director releases, an ACH file a shared-service team approves under a quorum rule, a card or instant-payment instruction, a stablecoin transfer, and an agent-initiated proposal.

Regulatory, identity, and risk information is not collapsed into a universal statement that a person is “compliant.” Each sourced fact is represented as an Evidence Assertion \href{https://chainit.com/product/data-tokenization/}{VDT} with provenance, freshness, scope, and privacy controls, evaluated at the moment of reliance rather than inherited from onboarding. The BRE returns the predicate results; the organization's configuration of the ChainIT Workflow Engine composes them into an explicit disposition (e.g., Allow, Allow with Controls, Step-Up, Review, Hold, Reject, Prohibit, or Stale/Unknown) and the organization is responsible for that policy logic and for its use of the resulting disposition, subject to applicable law and contractual allocation among ChainIT, the organization, payment providers, evidence providers, and other participating parties. This preserves the difference between organizational authority, sourced evidence, institutional policy, and present execution eligibility. The evidence, policy-profile and disposition mechanics are described in full in the companion paper, ChainIT \href{https://chainit.com/white-papers/}{Provable Compliance}. \cite{provable-compliance}

An agent is a subject of authority, not a source of it. A verified principal who holds authority under an ARP may grant a registered agent a narrower authority to act on their behalf, bounded by purpose, counterparty, transaction class, rail, per-transaction and aggregate value, and effective period. That grant is recorded in an append-only Agent Authority \href{https://chainit.com/product/data-tokenization/}{VDT}: provable evidence of what authority was granted, by whom, and within what bounds. The record is immutable, while present validity resolves at the reliance event against current ARP status, so it is evidence of a grant rather than a bearer credential. Within the granted band the agent acts with no human in the loop; crossing the boundary (a new payee, a changed destination, a higher value, an irreversible rail) requires biometric exact-digest approval by a person with independent authority for that transaction. The Agent Subject Profile adds only the identity and containment controls a non-human subject requires (a versioned Agent Registration VDT, an attested workload key authorizing an ephemeral session key, and the sovereignty boundary that keeps final execution authority outside the model); it adds no authority concept of its own.

\begin{greenbox}{Core proposition}
Before value moves or another consequential action executes, ChainIT is designed to establish: who or what is acting (a verified person, a group under quorum, an organization, or a permitted agent); whose authority supports the action; whether current evidence and policy permit that authority to be exercised now; what exact transaction is authorized; whether capacity is reserved; and what ultimately occurred.
\end{greenbox}

\section{The Problem: Authentication Is Not Portable Authority}\label{sec:problem}

Most payment infrastructure does more than merely validate a password or key. Banks and enterprises already maintain internal entitlements, approval limits, maker-checker controls, device authentication, sanctions screening, and fraud scoring. The missing capability is a portable, transaction-bound chain of authority that can be independently consumed across an organization, its bank, a processor, a marketplace, an agent runtime, and a counterparty.

The material question is therefore not only “Is this signature valid?” It is “Is this identified person or attested workload acting under a current organizational delegation, within the applicable evidence and policy state, against available capacity, on the same transaction that was evaluated and approved?” Without that answer, organizations either block agent-initiated activity or permit it through controls that are difficult to verify outside the system that created them.

The question is the same whether the initiating subject is a person, an organization initiating through a scheduled workflow, or an autonomous agent, and whether value moves over ACH, wire, card, an instant-payment rail, or a blockchain. Agent initiation raises the stakes because no human is present to notice an unfamiliar payee or a changed bank account. The control gap it exposes, however, is the ordinary fiat gap: authority, evidence, approval, and capacity are each verified inside systems the relying party cannot inspect, and none of them is cryptographically bound to the exact instruction that finally reaches the rail.

\begin{greenbox}{}
\begin{itemize}
  \item Authentication proves control of an accepted authenticator; it does not prove organizational power or transaction scope.
  \item A valid organizational delegation may exist while its exercise is temporarily held for evidence refresh, enhanced review, or step-up.
  \item A session may be valid while a particular payment is out of scope, over budget, directed to a new destination, or prohibited.
  \item A human approval is not sufficient unless it is bound to the exact transaction ultimately presented to the rail.
  \item A per-transaction limit is not enforceable under concurrency unless capacity is reserved atomically before commit authority is issued.
  \item A signed credential is not proof of current non-revocation unless live status or a sufficiently fresh signed status snapshot supports the decision.
\end{itemize}
\end{greenbox}
\vspace{19.8bp}\kern0bp%

\section{The ChainIT Architecture: Identity, Authority, Evidence, and Execution}\label{sec:architecture}

The architecture separates durable identity, organizational authority, sourced evidence, session eligibility, transaction policy, operational credentials, payment execution, and outcome evidence. This separation prevents one opaque credential from becoming a substitute for every decision and allows each relying party to consume only the facts needed for its purpose.

The architecture is organized in two layers. The \href{https://chainit.com/insights/chainit-launches-provable-authority-as-u-s-generative-ai-enabled-fraud-losses-are-projected-to-reach-40-billion-by-2027/}{Authority Protocol} is the general layer: it resolves whether an identified subject may perform a specific proposed action at the moment of reliance, and it is neither subject-specific nor rail-specific. The Agent Subject Profile is a narrower application layer that adds only the identity and containment controls a non-human subject requires: registration, workload attestation and key hierarchy, and the sovereignty boundary. It adds no authority concept of its own. An agent exercises delegated authority under the ARP exactly as a human delegate does, evidenced by an Agent Authority \href{https://chainit.com/product/data-tokenization/}{VDT}, which is a protocol-layer artifact. Every subsection below belongs to the Authority Protocol unless it is expressly identified as part of the Agent Subject Profile.

The subjects of authority are a verified individual, a group acting under a quorum or separation-of-duties rule, an organization acting through an authorized representative, and a permitted software agent. The ARP, Evidence \href{https://chainit.com/product/data-tokenization/}{VDTs}, Authority Policy Profile, BRE predicates, Workflow Engine disposition, canonical transaction digest, capacity reservation, Execution Authorization Credential, lifecycle VDTs, and \href{https://chainit.com/}{Valitorum} are the same across all four. What differs is how the subject is authenticated and registered, and what containment the subject requires.

Throughout this paper, the architecture is described in terms of the question each artifact answers, so that a relying party can consume only the facts its purpose requires.

\begin{figure}[H]
  \centering
  \includegraphics[width=435.58bp]{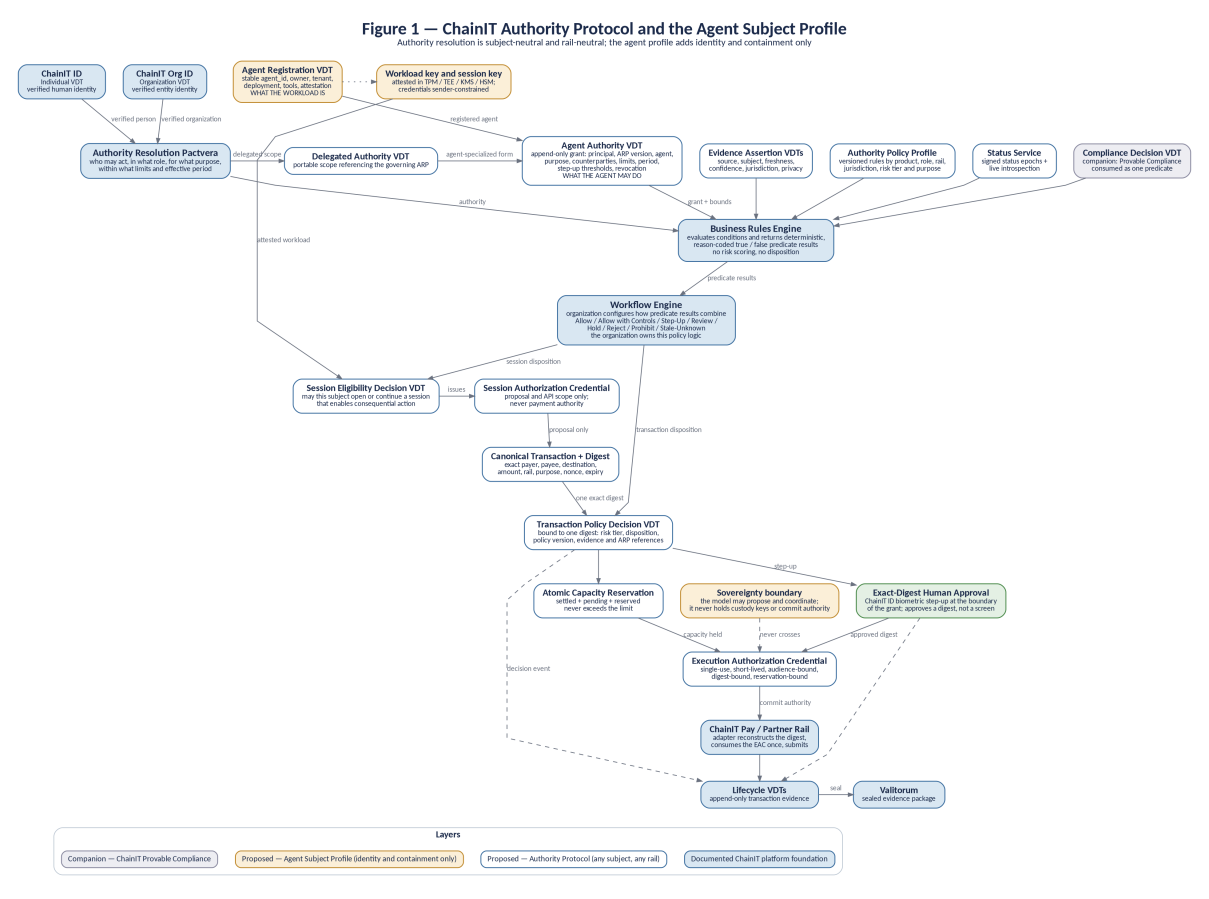}
  \caption{ChainIT \href{https://chainit.com/insights/chainit-launches-provable-authority-as-u-s-generative-ai-enabled-fraud-losses-are-projected-to-reach-40-billion-by-2027/}{Authority Protocol} and the Agent Subject Profile.}
  \label{fig:architecture}\vspace{-5.8bp}
\end{figure}

\subsection{ChainIT ID and the Individual VDT}\label{sec:chainit-id}

\href{https://chainit.com/product/identity-verification/}{ChainIT ID} provides the verified identity anchor for an individual. The supplied identity application materials describe authoritative-source verification, biometric and liveness controls, privacy-preserving and selective disclosure, portable identity storage, and verification without querying the originating government system on every use. Transaction-time reauthentication should therefore be described as comparison against a previously source-validated ChainIT ID under the applicable refresh and assurance policy, not automatically as a new government query for each approval. \cite{chainit-identity}

The factor combination — biometric, device, cryptographic wallet, PIN, location, or other controls — and the BeingID Level and Token Grade taxonomy (ChainIT\textquotesingle{}s internal classification of identity-proofing strength and credential quality) determine the assurance achieved. The required assurance is selected by risk and policy; a lower-assurance fallback cannot be described as equivalent to a higher-assurance approval merely because both occur through a ChainIT interface.

\subsection{ChainIT Org ID and the Organization VDT}\label{sec:org-id}

\href{https://chainit.com/product/business-verification/}{ChainIT Org ID} establishes the legal-entity identity anchor and may reference entity registration, ownership, officers, governance, authorized representatives, and related evidence. The Organization \href{https://chainit.com/product/data-tokenization/}{VDT} identifies the organization; it should not be used as a static container for every authorized person because authority changes independently. Authority belongs in a versioned ARP or a record derived from it. \cite{chainit-identity}-[\citen{marketplace}]

\subsection{Authority Resolution Pactvera, Delegated Authority, and the Agent Authority VDT}\label{sec:arp}

Authority Resolution \href{https://www.pactvera.com/}{Pactvera} (ARP) is the canonical, versioned organizational authority record: who may act, in what role, for what purpose, within what limits, during what effective period, under what quorum or separation-of-duties requirements, and subject to what amendment or revocation. Under the \href{https://chainit.com/insights/chainit-launches-provable-authority-as-u-s-generative-ai-enabled-fraud-losses-are-projected-to-reach-40-billion-by-2027/}{Authority Protocol}, a portable Delegated Authority \href{https://chainit.com/product/data-tokenization/}{VDT} may reference the governing ARP and expose only the scope needed by a delegate, an agent, or a relying party. The portable artifact does not replace the ARP, and its continued validity depends on current ARP status. \cite{pactvera,marketplace}

Authority is an organizational fact that can remain historically valid even when present exercise is suspended. An expired identity credential, an unresolved evidence alert, or a required enhanced review does not rewrite the historical delegation; it changes whether that authority may be exercised for the present context.

The division of labor with the companion paper follows from this. The ARP answers who may act, for what purpose, within what scope, and during what time. The Compliance Decision \href{https://chainit.com/product/data-tokenization/}{VDT} answers whether present evidence and policy permit that authority to be exercised now. The two are evaluated together, and neither substitutes for the other. A corporate officer remains an officer while a transaction is held for enhanced due diligence, and an expired identity credential reduces assurance without proving the person lost every corporate role. \cite{provable-compliance}

Delegation follows the same rule. A principal who holds authority under the ARP may grant a narrower authority to someone or something acting on their behalf, and the grant is bounded by purpose, counterparty, transaction class, rail, per-transaction and aggregate value, and effective period. Delegation to a registered agent is an instance of this, not an exception to it. An agent holds no authority of its own; it exercises a grant that a verified principal made under an ARP that permitted the delegation.

The Agent Authority \href{https://chainit.com/product/data-tokenization/}{VDT} is the append-only record of that grant, and the agent-specialized form of the Delegated Authority VDT. Each entry records the granting principal, the governing ARP and its version, the registered agent, the permitted purpose and counterparties, the transaction classes and rails, the per-transaction and aggregate limits, the effective period, the step-up thresholds, the redelegation rule, and the revocation conditions. Amendments, suspensions, and revocations create new linked entries rather than editing an earlier one, so a relying party, auditor, regulator, or reviewer can establish what authority the agent held at any past moment as well as what it holds now.

\runin{Immutable grant, resolved state.} The distinction matters. The record is immutable; present validity is not carried by the record. Authority changes (a principal lowers a limit, the governing ARP is amended, the delegation is revoked) and an artifact that asserted current authority on its own would be a portable bearer credential of exactly the kind this architecture is designed to prevent. The Agent Authority \href{https://chainit.com/product/data-tokenization/}{VDT} therefore proves what was granted and when; present validity resolves at the reliance event against current ARP status, revocation state, and the applicable freshness bound.

\runin{Necessary, not sufficient.} The Agent Authority \href{https://chainit.com/product/data-tokenization/}{VDT} establishes that authority was granted and within what bounds. It does not establish that the workload presenting it is the registered workload. That is attestation and sender-constrained proof. It does not establish that this transaction falls within scope right now. That is the policy evaluation at the reliance event. And it does not authorize the commit. That is the one-time Execution Authorization Credential bound to the exact canonical transaction digest. It is one required input to the execution predicate, not a credential that can be presented in place of the others.

\runin{The autonomous band and the step-up boundary.} A grant defines a band within which the agent may act without a human in the loop, and a boundary beyond which it may not. Within the band, an approved counterparty, an established transaction class, a value below the configured threshold, the agent proposes and the deterministic control plane decides, with no human interaction. Crossing the boundary — a new payee, a changed destination, a higher value, an irreversible rail, or a materially changed evidence state — requires biometric exact-digest approval by a person holding independent authority for that transaction, as described in Section~\ref{sec:human-approval}. Step-up is the boundary condition, not the ordinary case; a design that required human confirmation of every agent transaction would make the delegation meaningless. Human approval at the boundary satisfies a separate transaction-specific authority path under the ARP. It does not expand the standing grant, and the Agent Authority \href{https://chainit.com/product/data-tokenization/}{VDT} is not amended by it.

\subsection{Evidence Assertion VDTs and Context-Specific Decision VDTs}\label{sec:evidence}

Under the \href{https://chainit.com/insights/chainit-launches-provable-authority-as-u-s-generative-ai-enabled-fraud-losses-are-projected-to-reach-40-billion-by-2027/}{Authority Protocol} (for any subject, human or agent) the architecture does not use a universal Compliance \href{https://chainit.com/product/data-tokenization/}{VDT} asserting that a person or organization is broadly “in good standing.” Instead, each sourced fact is represented separately as an Evidence Assertion VDT. An assertion records the subject, source, source version, retrieval time, effective time, expiration, jurisdiction, match or confidence state, privacy classification, and any material limitation on use or disclosure.

\tealhead{Evidence Assertion VDTs}
Examples include identity-document status, entity status, ownership information, sanctions-screening result, PEP identification, license status, counterparty verification, beneficiary confirmation, device or location context, and unresolved monitoring alerts. Evidence is refreshed according to source-specific cadence, risk triggers, material changes, and policy. ChainIT need not re-query every external source for every transaction if a current assertion remains within its permitted freshness window. Each assertion is classified as authoritative-source, qualified third-party, or institution-generated evidence, and carries the freshness rule appropriate to that source class; the full taxonomy is set out in ChainIT \href{https://chainit.com/white-papers/}{Provable Compliance}. \cite{provable-compliance}

\tealhead{Session Eligibility Decision VDT}
A session-context decision answers whether a subject may open or continue a session that enables consequential action, and perform proposal or defined API activity within it. The artifact is protocol-general: a signed-in treasury session is subject to the same evaluation as a workload session. What differs is the input set — a person is admitted on identity assurance and role, a workload on registration, attestation, tool scope, and its Agent Authority \href{https://chainit.com/product/data-tokenization/}{VDT}. It references the ARP, applicable evidence, status, jurisdiction, policy version, disposition, reason codes, and expiration. Where the subject is a workload, it additionally references the Agent Registration VDT, Agent Authority VDT, attested deployment, and applicable tool scope. It is not transaction-specific and cannot authorize payment.

\tealhead{Transaction Policy Decision VDT}
A transaction-context decision is bound to one canonical transaction digest. It records the risk tier, disposition, applicable policy version, evidence references, ARP and subject references, required approvals or controls, expiration, and reason codes. It is created the same way whether the transaction was proposed by a person, an organization through a scheduled workflow, or an agent. It is newly derived for the proposed transaction and is not reusable as a permanent status attached to the person or organization.

\tealhead{Regulatory precision}
\begin{itemize}
  \item PEP identification is a risk factor evaluated through risk-based customer due diligence and monitoring; it is not itself an automatic disqualification. \cite{ffiec-pep}
  \item OFAC exposure is not limited to a name on the SDN list; direct and indirect ownership analysis and false-positive resolution may be required. \cite{ofac401}
  \item AML is an institutional program of monitoring, investigation, and reporting. An alert is evidence requiring evaluation, not a portable fact that a person is “AML invalid.”
  \item Where criminal-history information is used for employment or role qualification, relevance is contextual to the role, industry, jurisdiction, offense, and elapsed time; it is not a universal pass/fail condition. \cite{eeoc}
  \item A portable credential or shared ledger must not disclose a SAR or information that would reveal the existence or non-existence of a SAR. \cite{sar}
\end{itemize}

\begin{redbox}{SAR confidentiality constraint}
A relying party may receive a disposition such as “Hold: manual review required” without receiving the protected source or rationale that would disclose the existence or non-existence of a SAR. The evidence model must support restricted internal review and minimum counterparty disclosure.
\end{redbox}

\subsection{Business Rules Engine, Workflow Engine, and Execution Eligibility}\label{sec:bre}

The BRE is the deterministic predicate-evaluation point. For a session, it evaluates identity assurance, the ARP, evidence, status, jurisdiction, and scope, together with registration, attestation, tool scope, and session risk where the subject is a workload. For a transaction, it evaluates those inputs together with the exact transaction, current capacity, required approvals, status freshness, and rail-specific constraints. Each result is explicit and reason-coded rather than a black-box “approved” flag.

The BRE does not assign the disposition. It runs within the ChainIT Workflow Engine, where the organization configures which conditions apply, how predicate results combine, and which disposition states and downstream actions follow. ChainIT supplies the evidence, the predicate-evaluation engine, and the workflow tooling; the organization is responsible for the policy logic it configures and for its use of the resulting disposition, subject to applicable law and contractual allocation among ChainIT, the organization, payment providers, evidence providers, and other participating parties. Collapsing predicate evaluation into a single \textquotedbl{}all checks pass?\textquotedbl{} boolean increases false positives, overblocking, and disclosure risk.

At the authority layer, the general execution predicate is the one stated in ChainIT \href{https://chainit.com/white-papers/}{Provable Compliance}: MayExecute = IdentityAssuranceSatisfied AND AuthorityValidUnderARP AND ComplianceDecisionPermits AND TransactionWithinScope AND RequiredApprovalsSatisfied AND BudgetAndRailControlsSatisfied. The commit-time predicate below is that same test specialized to the payment boundary, where the exact instruction, the capacity reservation, and the one-time credential must also be verified. \cite{provable-compliance}

\begin{formulabox}{Commit-time execution predicate}
\formula{MayCommit = SubjectAuthenticated AND AuthorityValidUnderARP AND EvidenceAcceptable AND ComplianceDecisionPermits AND WorkflowDispositionIsAllow AND StatusFresh AND CapacityReserved AND RequiredApprovalsSatisfied AND DigestMatches AND EACValidAndUnused}
\formulanote{Every predicate is required. SubjectAuthenticated resolves to workload attestation where the subject is an agent and to \href{https://chainit.com/product/identity-verification/}{ChainIT ID} authentication at the required assurance level where the subject is a person. Allow with Controls or Step-Up must be completed and re-evaluated to Allow before EAC issuance or commit. Failure produces a controlled disposition, not silent fallback.}
\end{formulabox}

\subsection{Stable Agent Identity and Key Domains}\label{sec:agent-identity}

This subsection describes the Agent Subject Profile, which answers what the workload is rather than what it may do. Where the subject of authority is a verified person, the equivalent control is \href{https://chainit.com/product/identity-verification/}{ChainIT ID} authentication at the policy-required assurance level; no agent registration or workload key is involved, and no other part of the control flow changes. What the agent may do is a separate question, answered by the ARP and the Agent Authority \href{https://chainit.com/product/data-tokenization/}{VDT} in Section~\ref{sec:arp}.

Under the Agent Subject Profile, every autonomous workload is assigned a stable logical agent\_id and a versioned Agent Registration \href{https://chainit.com/product/data-tokenization/}{VDT} identifying its organization, responsible party or role, tenant, deployment, software or workload identity, permitted tools and APIs, risk class, attestation method, registered public key, status reference, and version history.

The key hierarchy is explicit. A rotatable workload key is generated or protected in an approved security boundary such as a TPM, TEE, KMS, HSM, or equivalent environment. That workload key authorizes an ephemeral session key. The Session Authorization Credential, per-request proofs, and Execution Authorization Credential are sender-constrained to the session key. ChainIT issuer keys sign and attest credentials; they do not act as the agent. The stable agent\_id persists when workload or session keys rotate.

Sender-constrained proof reduces the value of a copied credential because an attacker must also control the bound key. It does not, by itself, authenticate the organization, establish authority, validate policy, or prevent malicious code already operating inside the legitimate workload from using that key. \cite{rfc9449}

\subsection{Session and Execution Credentials}\label{sec:credentials}

Every artifact below is protocol-general. A session credential is never payment authority, and a session eligibility decision is never transaction authority, whether the session belongs to a workload or to a signed-in person; only a one-time credential bound to an exact canonical transaction digest may commit value. What varies by subject class is the input set each artifact evaluates, not the artifact itself. The agent-specific inputs (Agent Registration \href{https://chainit.com/product/data-tokenization/}{VDT} and the attested key hierarchy) are described in Sections~\ref{sec:agent-identity} and~\ref{sec:agent-auth}; the agent's authority to act at all is evidenced by the Agent Authority VDT in Section~\ref{sec:arp}.

\tablestyle{8bp}{10.85bp}{12.9bp}{6.9bp}\nopagebreak\begin{longtable}{|B{133.2bp}|M{183.4bp}|M{118.6bp}|}
\hline
\headrow\thc{Artifact} & \thc{Purpose} & \thc{Critical limitation}\\\hline
\endfirsthead
\hline
\headrow\thc{Artifact} & \thc{Purpose} & \thc{Critical limitation}\\\hline
\endhead
Session Eligibility Decision \href{https://chainit.com/product/data-tokenization/}{VDT} & Records that the authenticated subject may open or continue a session under the current context. & Not bound to an exact payment and cannot authorize value movement.\\\hline
\altrow Session Authorization Credential (SAC) & Short-lived, sender-constrained credential permitting proposal, evidence retrieval, and defined API activity. & Rail adapters must reject it as payment authority.\\\hline
Transaction Policy Decision \href{https://chainit.com/product/data-tokenization/}{VDT} & Records the risk tier, disposition, policy version, evidence state, controls, and exact transaction digest. & Only a final Allow disposition—after required controls and step-up are completed and revalidated—may support EAC issuance.\\\hline
\altrow Execution Authorization Credential (EAC) & Single-use, short-lived credential authorizing one exact transaction for one audience after reservation and approvals. & Bound to the session key, digest, reservation, status epoch, and expiration; consumed atomically.\\\hline
Execution Authorization Lifecycle \href{https://chainit.com/product/data-tokenization/}{VDT}(s) & Permanent evidence of EAC issuance and terminal handling (consumed, expired, canceled, or rejected), using one terminal record or separate linked issuance and disposition events. & Must not contain a reusable credential or private key; an earlier immutable event is never updated.\\\hline
\end{longtable}

\subsection{Agent Authentication and Session Bootstrap}\label{sec:agent-auth}

This subsection also describes the Agent Subject Profile. The corresponding step for a human subject is \href{https://chainit.com/product/identity-verification/}{ChainIT ID} authentication at the assurance level the Authority Policy Profile requires for the session. Both paths then converge on the same session-eligibility evaluation.

Before issuing a session, ChainIT must authenticate the registered workload. The workload presents its Agent Registration \href{https://chainit.com/product/data-tokenization/}{VDT} reference, attested workload key, approved deployment evidence, and proof of possession. The workload key then authorizes an ephemeral session key. The Session Eligibility Decision VDT is created for that context, and the SAC is sender-constrained to the session key.

A persistent bootstrap credential or a one-time enrollment credential may be used as an additional operational pattern, but neither is sufficient by itself. A persistent credential must be cryptographically bound to the registered workload. A one-time credential must be delivered through a channel authenticated to the correct workload, such as mutual Transport Layer Security (mTLS), private-key authentication, cloud workload identity, or another approved control.

\subsection{Pactvera, ChainIT Pay, Lifecycle VDTs, and Valitorum}\label{sec:pactvera}

\href{https://www.pactvera.com/}{Pactvera} and the \href{https://chainit.com/}{Valitorum} Service provide the transaction logic and final evidence. \href{https://chainit.com/product/digital-pay/}{ChainIT Pay}, or a compatible partner rail adapter, serves as the EAC enforcement boundary where the applicable \href{https://chainit.com/insights/chainit-launches-provable-authority-as-u-s-generative-ai-enabled-fraud-losses-are-projected-to-reach-40-billion-by-2027/}{Authority Protocol} profile is deployed, and each material state transition is recorded as a separate append-only lifecycle \href{https://chainit.com/product/data-tokenization/}{VDT}. This applies to every subject class; the enforcement point does not change because the proposal originated with an agent rather than a person. The Valitorum Service seals the completed or otherwise terminal evidence graph. \cite{pactvera}-[\citen{pat940}][\citen{chainitpay}]

\subsection{Relationship to Provable Compliance}\label{sec:compliance}

\href{https://chainit.com/insights/chainit-launches-provable-authority-as-u-s-generative-ai-enabled-fraud-losses-are-projected-to-reach-40-billion-by-2027/}{Provable Authority} and \href{https://chainit.com/white-papers/}{Provable Compliance} answer adjacent questions and are designed to be evaluated together. The ARP answers who may act, for what purpose, within what scope, and during what time. The Compliance Decision \href{https://chainit.com/product/data-tokenization/}{VDT} answers whether present evidence and policy permit that authority to be exercised now, for this product, counterparty, rail, jurisdiction, and risk tier. Neither substitutes for the other, and neither rewrites the other. \cite{provable-compliance}

The practical consequence is a division of effect. A compliance disposition gates the present exercise of authority; it does not amend, suspend, or delete the ARP. A newly identified politically exposed person retains her corporate role while her transactions receive enhanced review. An expired driver's license reduces identity assurance without proving the person ceased to hold office. Only a confirmed mandatory prohibition, or another expressly defined policy condition, produces broader blocking. Conversely, a favorable compliance disposition confers no authority: a fully evidenced, unrestricted subject still cannot execute a transaction the ARP does not permit.

Both papers use the same platform mechanics. Evidence is captured as sourced, time-bounded \href{https://chainit.com/product/data-tokenization/}{VDTs}; a versioned policy profile selects the applicable rules; the BRE returns deterministic predicate results; and the organization's configuration of the ChainIT Workflow Engine composes those results into a disposition the organization defines and owns. The \href{https://chainit.com/insights/chainit-launches-provable-authority-as-u-s-generative-ai-enabled-fraud-losses-are-projected-to-reach-40-billion-by-2027/}{Authority Protocol} consumes the compliance result as a single predicate, Compliance Decision Permits, and contributes the authority resolution, transaction digest, capacity reservation, exact-digest approval, execution credential, and sealed outcome evidence that a compliance decision alone does not provide.

\section{Verification at Runtime: Status, Risk Tier, and Disposition}\label{sec:runtime}

A signature and expiration establish that a credential was issued, has not been altered, and is within its stated validity period. They do not establish that the credential, agent registration, ARP, person, organization, or supporting authority has not been revoked since issuance. Current non-revocation therefore requires live status or a signed status artifact within an approved freshness bound.

Freshness is evaluated at the reliance event, not on a universal schedule. The reliance events for this protocol are session issuance or renewal, a transaction proposal or payment initiation, a human step-up approval, a scheduled review set by source volatility or relationship risk, and a source-change or material risk event such as a sanctions-list update, a license change, an ownership change, a policy update, or an ARP amendment. Each required item carries its own maximum age, derived from the source, the evidence type, the transaction risk, the product, the jurisdiction, and the reliance event itself — the same source-specific freshness rule stated in ChainIT \href{https://chainit.com/white-papers/}{Provable Compliance}. Reliance events are identical in kind whether the subject is a person or an agent. \cite{provable-compliance}

\begin{formulabox}{Bounded offline freshness}
\formula{StatusFresh = (CurrentTime - StatusSnapshot.IssuedAt) <= min(RiskTierStatusLimit, AuthorityPolicyStatusLimit)}
\formula{EvidenceFresh = AND, for each Evidence Assertion \href{https://chainit.com/product/data-tokenization/}{VDT} i: (CurrentTime - Evidence\_i.EffectiveTime) <= SourceAndPolicyLimit\_i}
\formulanote{StatusFresh tests whether the cached non-revocation snapshot for a credential, ARP, or registration is still within its approved bound. EvidenceFresh separately tests each underlying Evidence Assertion \href{https://chainit.com/product/data-tokenization/}{VDT} (sanctions, identity, licensing, beneficial-ownership, and similar sourced facts) against its own source- and policy-specific limit. A current status snapshot does not by itself establish that this underlying evidence is current. Commit requires both StatusFresh and EvidenceFresh; EvidenceFresh is one of the conditions within the broader EvidenceAcceptable test applied at commit (Section~\ref{sec:bre}). If either test fails, or the risk tier requires live status, the verifier must introspect or re-retrieve rather than accepting the cached value.}
\end{formulabox}

The status service publishes signed status epochs or compact status snapshots that verifiers can cache. A compatible compact status structure may follow established credential-status patterns, while a live service may follow introspection semantics adapted to ChainIT authority and transaction context. \cite{rfc7662,bsl}

\subsection{Risk Tier Selects Verification Rigor}\label{sec:risk-tier}
\vspace{-3.2bp}
\tablestyle{8bp}{10.85bp}{12.9bp}{6.9bp}\nopagebreak\begin{longtable}{|B{79.2bp}|M{205.2bp}|M{151.2bp}|}
\hline
\headrow\thc{Risk tier} & \thc{Illustrative condition} & \thc{Minimum verification mode}\\\hline
\endfirsthead
\hline
\headrow\thc{Risk tier} & \thc{Illustrative condition} & \thc{Minimum verification mode}\\\hline
\endhead
Tier 1 - Low & Known payee, low value, reversible flow, normal device and behavior. & Signed status snapshot within the longest approved bound; routine authentication and monitoring.\\\hline
\altrow Tier 2 - Standard & Normal vendor payment within established limits and ordinary risk. & Tighter status bound; live status when evidence changed or snapshot approaches its limit.\\\hline
Tier 3 - High & New payee, changed destination, high amount, cross-border, irreversible rail, anomalous device, or critical evidence change. & Live status, stronger authentication, step-up or quorum as required; no stale offline acceptance.\\\hline
\end{longtable}

\subsection{Policy Disposition Selects the Outcome}\label{sec:disposition}

The table below illustrates one disposition set an organization might configure in the ChainIT Workflow Engine using BRE predicate results as inputs. These are not decisions ChainIT renders on the organization's behalf. An organization may define a narrower or broader set to match its own policies and regulatory obligations and should align that set with the disposition states used in its Compliance Policy Profile.

\tablestyle{8.5bp}{11.5bp}{13.35bp}{7.1bp}\nopagebreak\begin{longtable}{|B{100.7bp}|D{345.5bp}|}
\hline
\headrow\thc{Disposition} & \thc{Meaning}\\\hline
\endfirsthead
\hline
\headrow\thc{Disposition} & \thc{Meaning}\\\hline
\endhead
Allow & All required predicates are satisfied; proceed to the next controlled stage. Commit still requires a valid reservation, EAC, and adapter checks.\\\hline
\altrow Allow with Controls & Proceed only after the specified additional control is completed and the transaction is re-evaluated to Allow.\\\hline
Step-Up & Additional human authentication, approval, quorum, or role validation is required.\\\hline
\altrow Review & A potential match, elevated risk, conflicting evidence, or a contextual judgment requires an authorized reviewer before the request can be dispositioned. A case is opened or referenced; reviewer authority and confidentiality are preserved.\\\hline
Hold & Evidence is incomplete, stale, ambiguous, or under review; no execution pending manual resolution.\\\hline
\altrow Reject & The request lacks valid authority, uses a revoked or expired credential, is outside scope, is malformed, or otherwise fails policy; execution is denied but not necessarily legally prohibited.\\\hline
Prohibit & A mandatory legal or sanctions restriction, or an authorized emergency kill-switch rule, prevents execution.\\\hline
\altrow Stale / Unknown & The system cannot support a current decision within the required freshness and confidence bounds. Fail controlled — hold, review, or deny according to the risk profile — rather than assuming the prior result still holds.\\\hline
\end{longtable}

Risk and disposition are independent. A Tier 3 transaction may ultimately be allowed after live status and dual approval. A Tier 1 transaction may still be rejected because the ARP was revoked. An unresolved sanctions name match ordinarily produces Hold or Review; a confirmed applicable prohibition produces Prohibit.

\begin{bluebox}{Recorded status evidence}
Every decision should record the signed snapshot or introspection result, status epoch, evaluation time, maximum permitted staleness, and reason the selected verification mode was acceptable. “Verified offline” is therefore a bounded and auditable statement, not a claim of instant universal revocation.
\end{bluebox}

\section{Applying the Protocol to Payment Pre-Authorization}\label{sec:payment}

Before value moves, the control plane evaluates five distinct questions: who or what is acting; whose authority supports the action; whether the current evidence and policy permit that authority to be exercised; whether capacity is available and reserved; and whether the instruction presented to the rail is exactly the transaction evaluated and authorized.

The routine state sequence is: Propose -> Canonicalize -> Evaluate -> Verify Status -> Reserve -> Step-Up if Required -> Re-evaluate -> Issue EAC -> Validate and Consume -> Submit -> Record Outcome -> Reconcile -> Seal.

The disposition set in Section~\ref{sec:disposition} is illustrative of one execution configuration an organization might implement in the ChainIT Workflow Engine, not a set ChainIT dictates. For that illustrative set, the execution effect of each disposition is as follows.

\begin{itemize}
  \item Allow: continue to the next controlled stage.
  \item Allow with Controls: proceed only after the specified additional control is completed, then re-evaluate to Allow.
  \item Step-Up: complete the required human or quorum action, then re-evaluate to Allow.
  \item Review: pause execution pending authorized review; re-enter evaluation after a recorded result.
  \item Hold: pause execution while evidence, status, or another prerequisite remains unresolved.
  \item Stale/Unknown: obtain fresh status or evidence, or apply the organization\textquotesingle{}s configured fail-controlled outcome.
  \item Reject: terminate the current execution attempt.
  \item Prohibit: terminate execution and apply any required legal, sanctions, incident, or kill-switch process the organization has configured.
\end{itemize}

Only a final Allow disposition permits EAC issuance.

\begin{figure}[H]
  \centering
  \includegraphics[width=413.98bp]{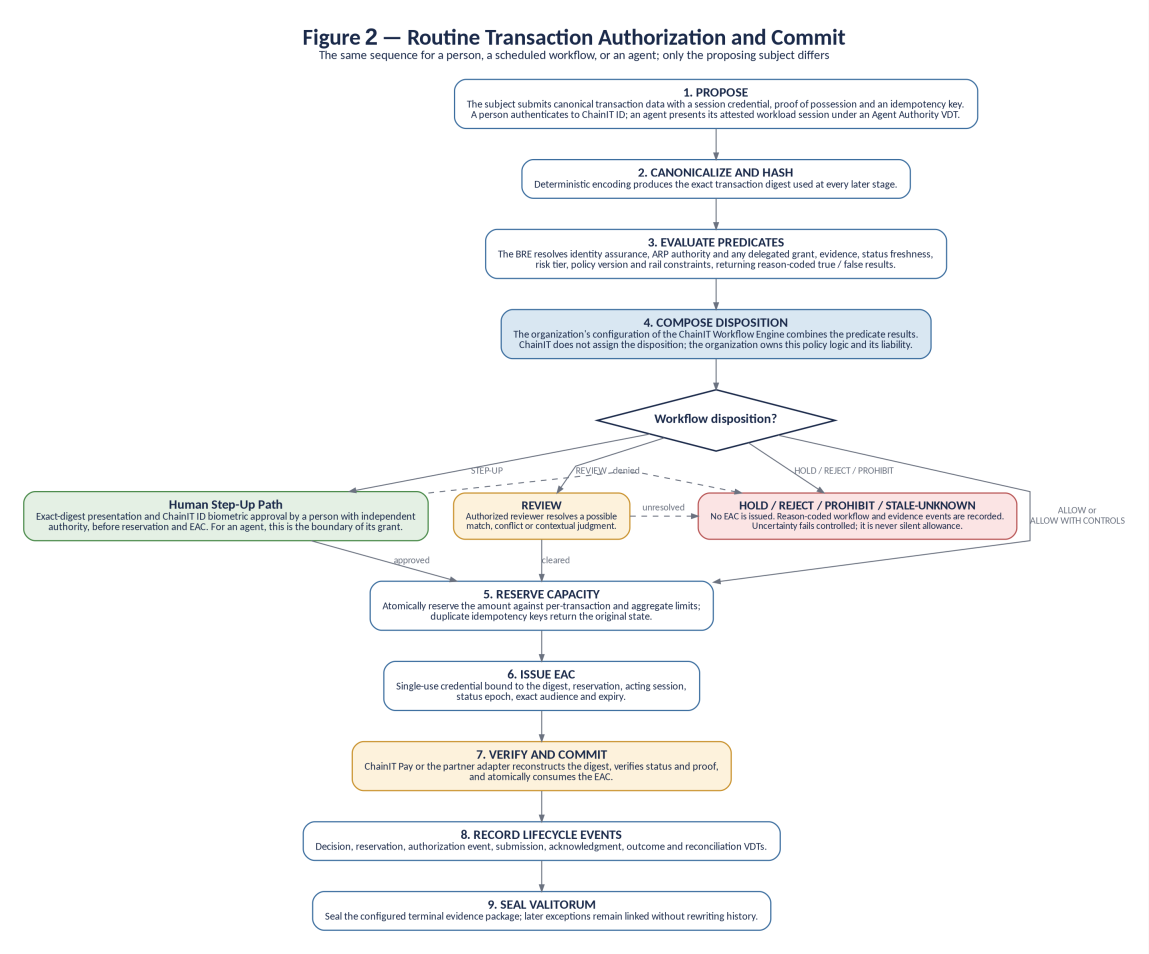}
  \caption{Routine Transaction Authorization and Commit.}
  \label{fig:commit}
  \fignote{The sequence is the same for a person, an organization initiating through a scheduled workflow, or an agent. A session credential permits proposal activity; only the one-time EAC authorizes commit of the exact canonical transaction.}
\end{figure}

\subsection{Subject Profiles: Human, Organizational, and Agent Initiation}\label{sec:subject-profiles}

The control sequence above is subject-neutral. What changes across subjects is the front of the sequence ( how the proposal originates and how the initiating subject is authenticated ) not the authority, evidence, policy, digest, reservation, execution, or outcome-evidence controls that follow it.

\runin{Human initiation.} A finance director prepares or approves a payment directly. She authenticates to \href{https://chainit.com/product/identity-verification/}{ChainIT ID} at the assurance level the Authority Policy Profile requires, her authority is resolved against the current ARP version, the canonical transaction object is constructed from the instruction actually being submitted, and capacity is reserved before an EAC is issued. The controls that matter most in ordinary fiat operations, that the approval is bound to the exact ACH or wire instruction transmitted, that a changed beneficiary account produces a different digest and therefore a new decision, and that two concurrent releases cannot draw on the same limit, are protocol controls, not agent controls.

\runin{Organizational or multi-party initiation.} A shared-service team, a scheduled disbursement run, or a treasury workflow initiates on behalf of the entity. Authority is resolved to the acting representatives and to the quorum or separation-of-duties rule the ARP imposes, and each required approver approves the same digest. The Human Approval \href{https://chainit.com/product/data-tokenization/}{VDTs} are separate, individually attributable events; a batch approval that is not bound to the individual canonical transactions does not satisfy the RequiredApprovalsSatisfied predicate.

\runin{Agent initiation.} A registered workload prepares the proposal under an Agent Authority \href{https://chainit.com/product/data-tokenization/}{VDT} granted by a principal with authority to delegate. The additional controls are those described in Sections~\ref{sec:agent-identity}, \ref{sec:agent-auth} and Section~\ref{sec:agent-profile}: registration, workload attestation, an ephemeral session key, and the sovereignty boundary. Authority resolution, session eligibility, evidence, policy, digest, reservation, approval and execution are unchanged. Within the granted band the sequence completes without human interaction; beyond it, the step-up path in Section~\ref{sec:human-approval} applies and a person with independent authority approves the exact digest. From the canonical transaction object onward, the sequence is identical to human initiation.

In every case the rail adapter enforces the same object: a one-time Execution Authorization Credential bound to a digest the adapter independently reconstructs. The rail profiles that follow describe how that enforcement is achieved on fiat rails and on-chain, and they apply to all three subject classes.

\subsection{Fiat Rail Profiles}\label{sec:fiat}

\goldlead[8.4bp]{Authority Protocol adapter with partner rail integration.}

This profile is the ordinary fiat case and does not presuppose an agent: the same preflight authorization, orchestration gate, or authorization call governs a wire a person releases. Existing ACH, wire, card, instant-payment, and bank interfaces do not universally accept arbitrary ChainIT credentials. The protocol therefore uses a rail-specific enforcement profile rather than assuming that the EAC is inserted into every native payment message. Reasonable models include a preflight authorization API, a \href{https://chainit.com/product/digital-pay/}{ChainIT Pay} orchestration gate, a bank or processor authorization call, a sidecar evidence object, or a compact authorization reference in supported metadata or ISO 20022 extensions.

The authoritative adapter must reconstruct the canonical transaction digest from the rail instruction, validate the EAC audience, session-key proof, status epoch, reservation, policy decision, approvals, expiration, and one-time-use state, and then consume the EAC through a recoverable commit path before or atomically with submission.

\subsection{Stablecoin and On-Chain Profiles}\label{sec:onchain}

\goldlead[8.4bp]{Authority Protocol adapter with custody, wallet, smart-contract, or network integration.}

For stablecoin or tokenized value, the control may be enforced by a policy-controlled custody wallet, smart-contract gate, account abstraction flow, relayer, paymaster, controlled signing service, or off-chain preflight tied to an approved transaction hash. The canonical object must bind, at minimum, the chain or network, token contract or asset, source wallet, destination wallet, amount, nonce, contract method or calldata hash, fee or gas policy, finality policy, and expiration.

\subsection{Outcome Events, Idempotency, and Unknown State}\label{sec:outcome-events}

Rail acknowledgments and outcomes must be authenticated and idempotent. A signed or mTLS-authenticated event should identify the ChainIT transaction and digest, rail transaction identifier, event type, sequence or ordering metadata, occurrence and receipt times, amount, currency or asset, reason code, prior-event reference, and idempotency key. Duplicate or out-of-order delivery must not create a duplicate payment or rewrite prior evidence.

A timeout after possible submission is an unknown state, not an instruction to retry blindly. The adapter must query or reconcile by idempotency key and rail reference before any new submission. This behavior is a core control, not a documentation detail.

\section{Exact-Digest Approval and Append-Only Lifecycle Evidence}\label{sec:digest-approval}

\subsection{Human Approval and Assurance}\label{sec:human-approval}

\goldlead{Biometric identity, exact-digest presentation, and subject-transaction binding operate together: the person approves the instruction, not a screen.}

When policy requires human action, the approver must possess independent organizational authority for the proposed transaction. Approval cannot cure a lack of authority. The approval experience presents the material transaction fields, risk warnings, relevant evidence or exceptions, and the approving role. The person then completes the authentication assurance required by policy and approves or denies the exact digest. This requirement is the same whether the proposal originated with another person, an organization\textquotesingle{}s scheduled workflow, or an agent: the approver approves a digest, not a screen, and approval of a different digest is not approval of this transaction. Where an agent proposal has crossed the step-up boundary set in its Agent Authority \href{https://chainit.com/product/data-tokenization/}{VDT}, this is the control that resolves it. The approval satisfies a separate transaction-specific authority path under the ARP; it does not widen the agent's standing grant, and the next transaction is evaluated against the same unchanged bounds.

Assurance is a property of the authentication method achieved - factors, phishing resistance, key protection, replay resistance, and identity binding - not simply whether the interaction occurred in an app, browser, or messaging channel. A browser flow may be high assurance; an SMS-only flow ordinarily is not. Higher-risk or irreversible transactions should require the approved high-assurance profile. \cite{nist63b}

\subsection{Canonical Transaction Digest: What You See Is What You Authorize}\label{sec:digest}

A canonical transaction object is encoded deterministically before hashing. The profile must define field names, schema version, decimal and currency representation, timestamp format, Unicode normalization, null handling, ordering, and amendment rules. RFC 8785 is one available model for deterministic JSON serialization. \cite{rfc8785}

\begin{formulabox}{Illustrative minimum digest inputs}
\formula{Digest = Hash(schema\_version | payer | source | payee | destination | amount | currency\_or\_asset | rail | purpose\_or\_invoice\_hash | fees | nonce | not\_before | expiration | rail\_specific\_parameters)}
\formulanote{Any material change requires a new object, digest, policy decision, reservation, approval, and EAC as applicable.}
\end{formulabox}

The same digest is bound to the Transaction Policy Decision \href{https://chainit.com/product/data-tokenization/}{VDT}, Reservation VDT, Human Approval VDT, EAC, Execution Authorization Lifecycle VDTs, rail instruction, and lifecycle evidence. The rail adapter independently reconstructs and compares it before submission.

\subsection{Operational EAC Versus Permanent Evidence}\label{sec:eac-evidence}

The EAC is a short-lived operational credential. It is not stored as a reusable ledger artifact. The permanent evidence layer uses Execution Authorization Lifecycle \href{https://chainit.com/product/data-tokenization/}{VDTs}. At minimum, a terminal event is written when the EAC is consumed, expires, is canceled, or is rejected; it records the EAC identifier or cryptographic hash, issuer, audience, transaction digest, reservation, status epoch, approvals, issuance and expiration times, validation result, terminal time, and linked submission reference. Where contemporaneous evidence at issuance is required, the system creates separate immutable EAC Issuance and EAC Consumption or Disposition VDTs. No earlier VDT is updated, and no private key or reusable credential is included.

\subsection{Append-Only Lifecycle VDTs and Valitorum}\label{sec:valitorum}

A sealed record is never modified. The transaction is represented by linked, immutable events: Payment Intent, Session or Transaction Decision, Reservation, Human Approval or Denial, Execution Authorization Issuance and Consumption or Disposition events, Rail Submission, Rail Acknowledgment, Settlement or other outcome, Reconciliation, and any exception event. Each event references the relevant parent or predecessor and preserves the source, time, signature, digest, and privacy classification.

\begin{figure}[H]
  \centering
  \includegraphics[width=399.59bp]{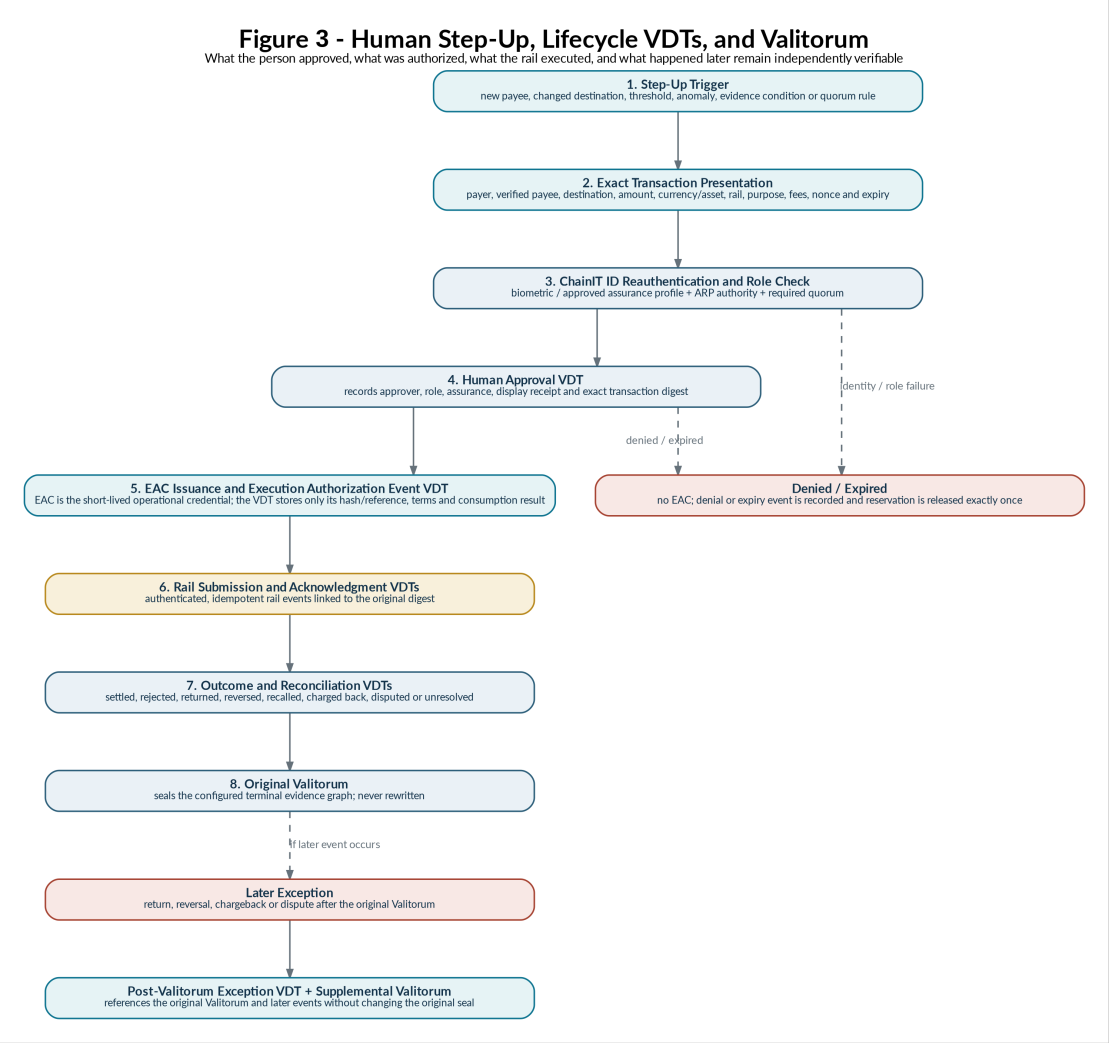}
  \caption{Human Step-Up, Lifecycle \href{https://chainit.com/product/data-tokenization/}{VDTs}, and \href{https://chainit.com/}{Valitorum}.}
  \label{fig:lifecycle}
  \fignote{The EAC is operational authority. Execution Authorization Lifecycle \href{https://chainit.com/product/data-tokenization/}{VDTs} are permanent evidence. The original \href{https://chainit.com/}{Valitorum} never changes.}
\end{figure}

The \href{https://chainit.com/}{Valitorum} Service seals the completed or otherwise terminal evidence graph under a defined completeness rule. If a later return, reversal, chargeback, recall, dispute, or post-settlement exception occurs, a new linked event is created. The original Valitorum remains immutable. The Valitorum Service may seal a Supplemental Valitorum over the later exception graph, referencing the original Valitorum so a relying party can verify both the original outcome and the subsequent change.

\href{https://www.pactvera.com/}{Pactvera} and the \href{https://chainit.com/}{Valitorum} Service provide structured evidence of identity, authority, terms, approvals, events, and outcomes. They do not by themselves guarantee enforceability, evidentiary admissibility, or allocation of liability in every jurisdiction or rail relationship.

\section{The Agent Subject Profile: Agentic Commerce and the Sovereignty Boundary}\label{sec:agent-profile}

Everything to this point describes the \href{https://chainit.com/insights/chainit-launches-provable-authority-as-u-s-generative-ai-enabled-fraud-losses-are-projected-to-reach-40-billion-by-2027/}{Authority Protocol}, which applies to any subject and any rail. This section describes the Agent Subject Profile: the identity and containment controls that exist only because the acting subject is autonomous. The agent's authority itself is not part of this profile. It is delegated under the ARP and evidenced by the Agent Authority \href{https://chainit.com/product/data-tokenization/}{VDT} described in Section~\ref{sec:arp}, in the same way authority is delegated to a human representative.

Agentic commerce requires two-sided verification. A relying party needs confidence that the workload presenting the request is the registered workload and that a verified, currently authorized human or organization stands behind the requested action. A legitimate agent identity can still act outside scope; a valid delegation can still be misused by an unregistered or compromised workload. Both sides must be proven together.

The architecture establishes a sovereignty boundary. An AI model or autonomous agent may reason, gather evidence, compare invoices, prepare terms, and propose an action. It does not receive unrestricted access to payment-account credentials, custody keys, or final execution authority. A deterministic control plane resolves authority, evidence, policy, status, capacity, and approval before issuing the one-time EAC that the rail adapter enforces. Section~\ref{sec:scenario-agent} works through this boundary end to end: an agent proposing routine invoices within its granted band, and the same agent\textquotesingle{}s proposal for a new, higher-value vendor crossing into the biometric exact-digest approval path described in Section~\ref{sec:human-approval}.

A verifier can trust a bounded set of ChainIT issuer keys and status services rather than maintaining every agent key independently. The cryptographic work per verification remains approximately constant, although total operational cost scales with transaction volume, evidence refresh, status distribution, support, and partner processing. Comparative economics therefore depend on deployment profile.

\section{Design Principles, Credential Lifetimes, and Residual Risk}\label{sec:principles}

\subsection{Zero-Trust Design Principles}\label{sec:zero-trust}

\tealhead{No implicit trust from prior success}
A prior identity proof, session, policy decision, or approval is not assumed to remain valid beyond its scope and freshness. The system reevaluates the predicates required for the present context.

\tealhead{Least privilege and intersection of constraints}
Effective authority is the intersection of human, organization, ARP, agent, counterparty, jurisdiction, policy, budget, and time constraints - never their union.

\tealhead{Separate identity, authority, evidence, policy, and execution}
Each artifact answers a defined question and can be refreshed, revoked, or disclosed independently.

\tealhead{Authority is separate from evidence and policy}
The ARP records organizational power and delegation. Evidence and compliance state govern whether that power may be exercised now. A risk indicator does not erase a valid delegation, and a generous delegation does not override a mandatory prohibition.

\tealhead{Organization-owned policy logic}
The BRE returns predicate results; the organization's configuration of the ChainIT Workflow Engine assigns the disposition. ChainIT supplies the evidence, the predicate engine, and the workflow tooling. The policy logic remains the organization's. The organization is responsible for its use of the resulting disposition, subject to applicable law and contractual allocation among ChainIT, the organization, payment providers, evidence providers, and other participating parties.

\tealhead{Sender-constrained operational credentials}
SACs and EACs require proof from the bound session key. Copied credentials are insufficient without the key. \cite{rfc9449}

\tealhead{Bounded status freshness}
Offline verification is accepted only within the risk-tier limit; live introspection is required when stale or high risk. \cite{rfc7662,bsl}

\tealhead{Atomic reservation and idempotency}
Capacity is reserved before EAC issuance; retries return the authoritative prior result rather than committing a second payment.

\tealhead{Fail controlled}
Invalid signature, unknown key, stale status, digest mismatch, unavailable required evidence, or ambiguous rail state produces Hold, Reject, Prohibit, or reconciliation - not silent acceptance.

\tealhead{Privacy by minimum disclosure}
Counterparties receive the decision and proof needed for their purpose, not raw biometrics, government records, SAR information, or unrelated identity details.

\subsection{Credential and Record Lifetimes}\label{sec:lifetimes}

\href{https://chainit.com/insights/chainit-launches-provable-authority-as-u-s-generative-ai-enabled-fraud-losses-are-projected-to-reach-40-billion-by-2027/}{Authority Protocol} artifacts apply to every subject class. Only the Agent Registration \href{https://chainit.com/product/data-tokenization/}{VDT}, the workload key, and the ephemeral session key are Agent Subject Profile artifacts. The Delegated Authority VDT and its agent-specialized form the Agent Authority VDT, the Evidence Assertion VDT, the Session Eligibility Decision VDT, the SAC, the Transaction Policy Decision VDT, capacity reservation, EAC, Execution Authorization Lifecycle VDTs, and Supplemental \href{https://chainit.com/}{Valitorum} are protocol-general; their inputs vary by subject class.

\tealhead{Identity, Authority, and Session Artifacts}
\tablestyle{7.5bp}{10.15bp}{12.35bp}{6.8bp}\nopagebreak\begin{longtable}{|B{125.9bp}|Y{71.6bp}|M{132.9bp}|M{93.4bp}|}
\hline
\headrow\thc{Credential or record} & \thc{Layer} & \thc{Typical lifetime or state} & \thc{Control boundary}\\\hline
\endfirsthead
\hline
\headrow\thc{Credential or record} & \thc{Layer} & \thc{Typical lifetime or state} & \thc{Control boundary}\\\hline
\endhead
Individual \href{https://chainit.com/product/data-tokenization/}{VDT} / \href{https://chainit.com/product/identity-verification/}{ChainIT ID} & Platform foundation & Durable, versioned identity anchor; subject to refresh, suspension, revocation, and recovery. & User-controlled credential architecture and ChainIT verification services, as implemented.\\\hline
\altrow Organization \href{https://chainit.com/product/data-tokenization/}{VDT} / \href{https://chainit.com/product/business-verification/}{ChainIT Org ID} & Platform foundation & Durable, versioned entity anchor. & ChainIT Org ID and governed organization records.\\\hline
ARP & Platform foundation & Versioned; effective until amended, superseded, expired, or revoked. & Organization authority domain.\\\hline
\altrow Delegated Authority \href{https://chainit.com/product/data-tokenization/}{VDT} & \href{https://chainit.com/insights/chainit-launches-provable-authority-as-u-s-generative-ai-enabled-fraud-losses-are-projected-to-reach-40-billion-by-2027/}{Authority Protocol} & Bounded period; validity depends on governing ARP status. & Portable reference held or presented by authorized component.\\\hline
Agent Authority \href{https://chainit.com/product/data-tokenization/}{VDT} & \href{https://chainit.com/insights/chainit-launches-provable-authority-as-u-s-generative-ai-enabled-fraud-losses-are-projected-to-reach-40-billion-by-2027/}{Authority Protocol} & Append-only grant record; each amendment, suspension, or revocation is a new linked entry. Present validity resolves at the reliance event against current ARP status. & Granting principal under the ARP; ChainIT evidence ledger.\\\hline
\altrow Evidence Assertion \href{https://chainit.com/product/data-tokenization/}{VDT} & \href{https://chainit.com/insights/chainit-launches-provable-authority-as-u-s-generative-ai-enabled-fraud-losses-are-projected-to-reach-40-billion-by-2027/}{Authority Protocol} & Source-specific effective and expiration periods. & Evidence service; selective disclosure.\\\hline
Session Eligibility Decision \href{https://chainit.com/product/data-tokenization/}{VDT} & \href{https://chainit.com/insights/chainit-launches-provable-authority-as-u-s-generative-ai-enabled-fraud-losses-are-projected-to-reach-40-billion-by-2027/}{Authority Protocol} & One session context; expires and is re-derived. & Authority gateway / BRE.\\\hline
\altrow Agent Registration \href{https://chainit.com/product/data-tokenization/}{VDT} & Agent Subject Profile & Versioned across deployments and key rotations. & Agent registry.\\\hline
Workload key & Agent Subject Profile & Rotatable; protected in approved TPM/TEE/KMS/HSM or equivalent boundary. & Registered workload security boundary.\\\hline
\altrow Ephemeral session key & Agent Subject Profile & One session or shorter. & Authorized by workload key; used for request proofs.\\\hline
SAC & \href{https://chainit.com/insights/chainit-launches-provable-authority-as-u-s-generative-ai-enabled-fraud-losses-are-projected-to-reach-40-billion-by-2027/}{Authority Protocol} & Minutes to hours; proposal and API scope only. & Session and authority gateway.\\\hline
\end{longtable}

\tealhead{Transaction, Execution, and Evidence Artifacts}
\tablestyle{7.5bp}{10.15bp}{12.35bp}{6.8bp}\nopagebreak\begin{longtable}{|B{125.9bp}|Y{71.6bp}|M{132.9bp}|M{93.4bp}|}
\hline
\headrow\thc{Credential or record} & \thc{Layer} & \thc{Typical lifetime or state} & \thc{Control boundary}\\\hline
\endfirsthead
\hline
\headrow\thc{Credential or record} & \thc{Layer} & \thc{Typical lifetime or state} & \thc{Control boundary}\\\hline
\endhead
Transaction Policy Decision \href{https://chainit.com/product/data-tokenization/}{VDT} & \href{https://chainit.com/insights/chainit-launches-provable-authority-as-u-s-generative-ai-enabled-fraud-losses-are-projected-to-reach-40-billion-by-2027/}{Authority Protocol} & One exact transaction and limited evaluation window. & BRE / transaction control plane.\\\hline
\altrow Capacity reservation & \href{https://chainit.com/insights/chainit-launches-provable-authority-as-u-s-generative-ai-enabled-fraud-losses-are-projected-to-reach-40-billion-by-2027/}{Authority Protocol} & Until commit, release, cancellation, or expiration. & Reservation ledger.\\\hline
EAC & \href{https://chainit.com/insights/chainit-launches-provable-authority-as-u-s-generative-ai-enabled-fraud-losses-are-projected-to-reach-40-billion-by-2027/}{Authority Protocol} & Seconds to minutes; single-use. & Issued to session; consumed by rail adapter.\\\hline
\altrow Execution Authorization Lifecycle \href{https://chainit.com/product/data-tokenization/}{VDT}(s) & \href{https://chainit.com/insights/chainit-launches-provable-authority-as-u-s-generative-ai-enabled-fraud-losses-are-projected-to-reach-40-billion-by-2027/}{Authority Protocol} & Permanent immutable evidence events for issuance and terminal handling. & ChainIT evidence ledger.\\\hline
Lifecycle \href{https://chainit.com/product/data-tokenization/}{VDTs} & Platform foundation & Permanent once written; new events represent later state. & \href{https://www.pactvera.com/}{Pactvera} / evidence ledger.\\\hline
\altrow \href{https://chainit.com/}{Valitorum} & Platform foundation & Permanent immutable seal over the configured terminal evidence graph. & Valitorum verification package.\\\hline
Supplemental \href{https://chainit.com/}{Valitorum} & \href{https://chainit.com/insights/chainit-launches-provable-authority-as-u-s-generative-ai-enabled-fraud-losses-are-projected-to-reach-40-billion-by-2027/}{Authority Protocol} & Permanent seal over a later exception graph referencing the original Valitorum. & Post-terminal exception verification package.\\\hline
\end{longtable}

\subsection{Residual Risk and Non-Goals}\label{sec:residual}

\href{https://chainit.com/insights/chainit-launches-provable-authority-as-u-s-generative-ai-enabled-fraud-losses-are-projected-to-reach-40-billion-by-2027/}{Provable Authority} is designed to prevent, limit, or expose specified classes of unauthorized execution. It does not eliminate:
\begin{itemize}
  \item A properly authorized person being deceived about the business purpose, beneficiary, or legitimacy of an invoice.
  \item Coercion, collusion, or malicious conduct by people who satisfy the formal authority and approval rules.
  \item Incorrect, delayed, or incomplete data from an authoritative source or evidence provider.
  \item Compromise of ChainIT issuer infrastructure, an HSM or KMS administrator, a payment partner, custody system, or payment rail.
  \item Compromise of a legitimate workload while it remains able to use its protected key.
  \item Residual exposure accepted by a verifier that uses a signed status snapshot up to the approved maximum staleness.
  \item Irreversible settlement after a valid but fraudulently induced approval.
\end{itemize}

\begin{redbox}{Claims discipline}
\href{https://chainit.com/insights/chainit-launches-provable-authority-as-u-s-generative-ai-enabled-fraud-losses-are-projected-to-reach-40-billion-by-2027/}{Provable Authority} is designed to reduce unauthorized, stale, replayed, substituted, duplicate, and out-of-scope transaction risk. It is not fraud-proof, non-spoofable, unstealable, tamper-proof, or instantaneously revoked everywhere, and it is not described as such. The residual risks above are stated so that a deploying organization can control for them.
\end{redbox}

\section{Illustrative Scenario}\label{sec:scenario}

Acme Corp verifies its organization through \href{https://chainit.com/product/business-verification/}{ChainIT Org ID}. Its governing body executes an ARP granting Jane, the finance director, personal authority to approve vendor payments up to \$100,000 per transaction, subject to a second authorized approver above \$25,000. The same ARP permits Jane to delegate a much narrower autonomous execution authority to a registered procurement agent: approved cloud-services vendors only, up to \$5,000 per transaction and \$25,000 per month, in U.S. dollars, with no redelegation. The ARP separately permits the agent to prepare and route higher-value or new-payee proposals for human decision, but not to execute them without the required authorized approvers.

\subsection{Human-initiated wire under the same ARP}\label{sec:scenario-human}

Jane initiates a \$60,000 wire to a long-standing supplier. No agent is involved. She authenticates to \href{https://chainit.com/product/identity-verification/}{ChainIT ID} at the high-assurance profile the Authority Policy Profile requires for an irreversible rail. The BRE evaluates her identity assurance, her authority under the current ARP version, payee and destination evidence, sanctions and entity evidence within their permitted freshness windows, and the second-approver requirement that applies above \$25,000. Acme's configuration of the ChainIT Workflow Engine composes those predicate results into a Step-Up disposition.

The canonical transaction object is built from the instruction that will actually be transmitted, including the beneficiary bank account. Capacity is provisionally reserved before the approval request is issued to Jane and the second approver, so that no concurrent request can consume the same capacity while approval is pending. After both authorize the same digest, policy and status are revalidated, the reservation is confirmed, and a one-time EAC is issued to the wire adapter, which independently reconstructs the digest before submission. Had the beneficiary account been changed after the first approval (the ordinary vendor-impersonation pattern) the digest would differ, the existing approvals would not apply, and a new decision, reservation, and approval cycle would be required. None of these controls depends on an agent being present.

\subsection{Agent-initiated payment within the delegated scope}\label{sec:scenario-agent}

The agent is registered under Acme's \href{https://chainit.com/product/business-verification/}{Org ID}. Jane's delegation is recorded as an Agent Authority \href{https://chainit.com/product/data-tokenization/}{VDT} referencing the governing ARP version, naming her as the granting principal, and fixing the approved-vendor scope, the \$5,000 and \$25,000 limits, the currency, the no-redelegation rule, and the thresholds above which biometric step-up is required. Its workload key is attested and authorizes an ephemeral session key. ChainIT creates a Session Eligibility Decision VDT and issues a sender-constrained SAC permitting invoice review and payment proposals. The agent does not receive unrestricted access to Acme's bank account, wallet, or rail credentials.

For a routine \$3,800 invoice to an established vendor, the agent creates the canonical transaction object and per-request proof. The BRE evaluates the applicable predicates against current status, evidence, and the agent\textquotesingle{}s authorized scope, and Acme\textquotesingle{}s configuration of the ChainIT Workflow Engine creates a Transaction Policy Decision \href{https://chainit.com/product/data-tokenization/}{VDT} and composes those predicate results into an Allow disposition. The reservation ledger atomically reserves \$3,800, and the authority gateway issues a one-time EAC bound to the session key, digest, reservation, status epoch, rail adapter, and expiration. The adapter reconstructs the digest, consumes the EAC once, submits the instruction, and records the authorization, submission, acknowledgment, settlement, and reconciliation events before the \href{https://chainit.com/}{Valitorum} Service seals the completed evidence graph.

Later, two \$4,000 proposals arrive simultaneously while only \$5,000 of monthly autonomous capacity remains. The reservation ledger serializes the competing requests. One reservation succeeds; the other receives an insufficient-capacity disposition. Both requests cannot read the same balance and both proceed.

The agent then prepares a \$50,000 proposal to a new vendor. The amount is outside the agent's autonomous execution scope but within the proposal-routing authority and Jane's personal approval authority. The Transaction Policy Decision \href{https://chainit.com/product/data-tokenization/}{VDT} requires live status, verified payee and destination evidence, a provisional capacity reservation, Jane's exact-digest approval, and the second authorized approver required above \$25,000. After both authorized people approve the same digest, the system revalidates policy and status, confirms the reservation, and may issue a one-time EAC. Human approval does not retroactively expand the agent's standing delegation; it satisfies a separate transaction-specific authority path under the ARP. The Agent Authority VDT is unchanged, and the next proposal is evaluated against the same \$5,000 and \$25,000 bounds.

\subsection{Exception handling under either subject profile}\label{sec:scenario-exception}

If a routine vendor later produces a possible sanctions name match, the transaction is held for resolution rather than automatically treated as prohibited. If the match is resolved as false, a new decision may allow execution. If an applicable prohibition is confirmed, the disposition is Prohibit and no EAC is issued. If the signed status snapshot for the vendor\textquotesingle{}s entity or sanctions evidence ages past its approved bound at the moment of reliance, the disposition resolves to Stale/Unknown rather than proceeding on the cached result; the verifier must obtain a live status introspection before the proposal can be re-evaluated to Allow. If a settled payment is later returned, the Settlement \href{https://chainit.com/product/data-tokenization/}{VDT} remains unchanged; a linked Return VDT and, when configured, a Supplemental \href{https://chainit.com/}{Valitorum} records the later event.

\section{Related Work}\label{sec:related}

\tealhead{Verifiable credentials and credential status}
The W3C Verifiable Credentials Data Model \cite{vcdm2} and Decentralized Identifiers \cite{did} standardize cryptographically verifiable claims about subjects, and the Bitstring Status List \cite{bsl} provides a privacy-preserving mechanism for publishing revocation and suspension status. \href{https://chainit.com/product/data-tokenization/}{Validated Data Tokens} play a comparable role for individual facts, but the \href{https://chainit.com/insights/chainit-launches-provable-authority-as-u-s-generative-ai-enabled-fraud-losses-are-projected-to-reach-40-billion-by-2027/}{Authority Protocol} deliberately avoids a single credential asserting that a subject is broadly in good standing: each sourced fact remains a separate Evidence Assertion VDT, and decisions are newly derived per reliance event rather than carried as standing status (Section~\ref{sec:evidence}). NIST SP~800-63B-4 \cite{nist63b} defines the authentication assurance on which the approval requirements of Section~\ref{sec:human-approval} rely.

\tealhead{Authorization, delegation, and capabilities}
OAuth~2.0 \cite{rfc6749} and its extensions are the dominant framework for delegated API authorization. Rich Authorization Requests \cite{rfc9396} allow fine-grained, transaction-specific authorization details, Token Exchange \cite{rfc8693} expresses delegation and impersonation chains, GNAP \cite{rfc9635} generalizes the grant negotiation model, and DPoP \cite{rfc9449} sender-constrains tokens to a client key; token introspection \cite{rfc7662} supplies live status. Capability-based security \cite{dennis1966}, SPKI authorization certificates \cite{rfc2693}, trust-management systems \cite{blaze1996}, and attenuable bearer tokens such as Macaroons \cite{macaroons} express delegated, narrowed authority. These mechanisms convey what a client or delegate is permitted to request; the \href{https://chainit.com/insights/chainit-launches-provable-authority-as-u-s-generative-ai-enabled-fraud-losses-are-projected-to-reach-40-billion-by-2027/}{Authority Protocol} builds on the same ideas but resolves authority against a versioned organizational record (the ARP) at the moment of reliance, keeps the grant record separate from present validity, and authorizes commit only through a single-use credential bound to one canonical transaction digest and an atomic capacity reservation.

\tealhead{Separation of duties and transaction binding}
Separation of duties and well-formed transactions are classical requirements of commercial integrity policy \cite{clarkwilson}, and role-based access control \cite{sandhu1996} is the standard way to express entitlements inside an enterprise. The difficulty of ensuring that a person approves the transaction actually executed, rather than what a screen displays, is known as the “what you see is what you sign” problem \cite{landrock1998}. The European strong-customer-authentication rules require dynamic linking of an authentication code to the specific amount and payee \cite{psd2rts}. The canonical transaction digest of Section~\ref{sec:digest} generalizes this binding to every material field, including rail-specific parameters, and extends it from the approval to the policy decision, reservation, execution credential, and lifecycle evidence.

\tealhead{Workload identity, attestation, and zero trust}
The Agent Subject Profile draws on workload identity frameworks such as SPIFFE \cite{spiffe}, the IETF Remote Attestation Procedures architecture \cite{rfc9334}, and the zero-trust principle that no implicit trust is granted on the basis of network location or prior success \cite{nist207}.

\tealhead{Identity and delegation for AI agents}
Recent work argues that AI agents require authenticated, auditable delegation from human principals, proposing extensions to OAuth~2.0 and OpenID Connect with agent-specific credentials \cite{south2025}, and that instances of AI systems should carry identifiers linking them to accountable parties \cite{chan2024ids}. Industry protocols for agent-initiated payments have also emerged. Google's Agent Payments Protocol expresses user authorization as signed mandates that are bound to transaction contents by hash and key confirmation, and returns receipts \cite{ap2}; the Agentic Commerce Protocol from OpenAI and Stripe standardizes agent checkout \cite{acp}; and Visa's Trusted Agent Protocol authenticates agents to merchants \cite{visatap}. Lan et al. \cite{lan2026} add runtime context binding and consume-once enforcement to AP2 to mitigate replay and context-binding failures under concurrency. The \href{https://chainit.com/insights/chainit-launches-provable-authority-as-u-s-generative-ai-enabled-fraud-losses-are-projected-to-reach-40-billion-by-2027/}{Authority Protocol} shares the goals of transaction binding and single use. It differs in where authority comes from and what is checked before commit: it treats an agent as one subject class among four, resolves the agent's authority against the same versioned organizational authority record (the ARP) that governs human delegates, evaluates sourced evidence and status freshness at each reliance event, and reserves capacity atomically against delegated per-transaction and aggregate limits before a single-use execution credential is issued.

\tealhead{Tamper-evident evidence}
Append-only lifecycle \href{https://chainit.com/product/data-tokenization/}{VDTs} and the sealed \href{https://chainit.com/}{Valitorum} relate to tamper-evident logging \cite{crosby2009} and transparency logs such as Certificate Transparency \cite{rfc9162}, which allow later verification that recorded history has not been rewritten.

\section{Conclusion}\label{sec:conclusion}

The control question in this paper is not specific to agents. It is whether an identified subject (a person, a group acting under quorum, an organization, or a permitted workload) is acting under a current organizational delegation, within the applicable evidence and policy state, against available capacity, on the same transaction that was evaluated and approved. That question often cannot be independently answered outside the systems that enforce the underlying authority, approval, and capacity controls, none of which is ordinarily bound to the instruction that finally reaches the rail. Agentic commerce does not create the question. It removes the human who would otherwise have noticed that the answer was wrong.

ChainIT \href{https://chainit.com/insights/chainit-launches-provable-authority-as-u-s-generative-ai-enabled-fraud-losses-are-projected-to-reach-40-billion-by-2027/}{Provable Authority} supplies a structured answer by connecting verified identity, organizational authority, sourced evidence, context-specific policy, transaction semantics, human approval, payment execution, and immutable outcome evidence. Applied to autonomous workloads, the Agent Subject Profile additionally keeps AI outside unrestricted private-key and final execution control while allowing agents to propose and coordinate real economic activity. Where the companion Compliance Decision \href{https://chainit.com/product/data-tokenization/}{VDT} supplies the evidence-and-policy predicate, the Authority Protocol supplies the authority resolution, transaction binding, capacity control, approval, execution, and sealed outcome evidence around it. \cite{provable-compliance} The result is not merely proof that a credential was presented. It is verifiable evidence of who or what acted, whose authority supported the action, what was authorized, which rules and evidence were applied, how capacity was controlled, and what ultimately occurred.

\clearpage
\unnumbered{References and Source Basis}

The architecture is grounded in the ChainIT patent and application materials and in the official technical and regulatory references listed below. External references are used for protocol design and regulatory framing, not as a substitute for deployment-specific legal or compliance review.

\begingroup
\renewcommand{\section}[2]{}%
\refsize

\endgroup


\begin{thebibliography}{99}
\setlength{\itemsep}{2bp}\setlength{\parskip}{0bp}\RaggedRight

\bibitem{chainit-identity} Portable Self-Sovereign Digital Identity with Immutable Trust \& Privacy-Preserving Verification, ChainIT application materials, including authoritative-source identity, biometric authentication, selective disclosure, portable verification, and credential update/revocation concepts; see also U.S. Patent Application Publication No. US 2025/0165575 A1, System for Creating, Storing, and Portably Utilizing Verified Digital Identities.

\bibitem{pactvera} \href{https://www.pactvera.com/}{Pactvera} Applications A and B, ChainIT application materials, including \href{https://chainit.com/product/identity-verification/}{ChainIT ID}, \href{https://chainit.com/product/business-verification/}{ChainIT Org ID}, Authority Resolution Pactvera, BRE-controlled execution, \href{https://chainit.com/product/data-tokenization/}{VDTs}, conditional value transfer, transaction validation, and \href{https://chainit.com/}{Valitorum}.

\bibitem{marketplace} System and Method for Managing Verified Digital Identities, Tokenized Assets, and Conditional Transactions in a Blockchain-Based Marketplace, ChainIT application materials, including \href{https://chainit.com/product/data-tokenization/}{Validated Data Token} terminology, ARP, BRE, \href{https://www.pactvera.com/}{Pactvera}, ZKP, and \href{https://chainit.com/}{Valitorum}.

\bibitem{pat772} U.S. Patent No. 11,216,772 B2, Use of Blockchain-Based Distributed Ledger and Smart Contracts for a Construction Project, including condition-driven electronic payment and immutable ledger records.

\bibitem{pat761} U.S. Patent No. 11,288,761 B2, Decentralized System for Verifying Participants to an Activity, including authenticated participants, devices, location, time, metadata, and immutable event records.

\bibitem{pat940} U.S. Patent No. 12,321,940 B2, System for Verification and Management for Digitally Cash Transactions, including payment records, transaction participants, amount, location, time, and persistent storage.

\bibitem{rfc9449} IETF RFC 9449, OAuth 2.0 Demonstrating Proof of Possession (DPoP), \url{https://www.rfc-editor.org/rfc/rfc9449.html}.

\bibitem{rfc7662} IETF RFC 7662, OAuth 2.0 Token Introspection, \url{https://www.rfc-editor.org/rfc/rfc7662.html}.

\bibitem{rfc8785} IETF RFC 8785, JSON Canonicalization Scheme (JCS), \url{https://www.rfc-editor.org/rfc/rfc8785.html}.

\bibitem{bsl} W3C Recommendation, Bitstring Status List v1.0: Privacy-Preserving Status Information for Verifiable Credentials (May 15, 2025), \url{https://www.w3.org/TR/vc-bitstring-status-list/}.

\bibitem{nist63b} NIST Special Publication 800-63B-4, Digital Identity Guidelines: Authentication and Authenticator Management (July 2025), \url{https://csrc.nist.gov/pubs/sp/800/63/b/4/final}.

\bibitem{ffiec-pep} FFIEC BSA/AML Examination Manual, Politically Exposed Persons, \url{https://bsaaml.ffiec.gov/manual/RisksAssociatedWithMoneyLaunderingAndTerroristFinancing/20}.

\bibitem{ofac401} U.S. Department of the Treasury, Office of Foreign Assets Control, FAQ 401 and the 50 Percent Rule, \url{https://ofac.treasury.gov/faqs/401}.

\bibitem{sar} 31 U.S.C. \S~5318(g)(2); Financial Crimes Enforcement Network Advisory FIN-2010-A014, Maintaining the Confidentiality of Suspicious Activity Reports, \url{https://www.fincen.gov/resources/advisories/fincen-advisory-fin-2010-a014}; and Financial Crimes Enforcement Network Advisory FIN-2012-A002, addressing confidentiality of the existence or non-existence of a SAR, \url{https://www.fincen.gov/resources/advisories/fincen-advisory-fin-2012-a002}.

\bibitem{eeoc} U.S. Equal Employment Opportunity Commission, Arrest and Conviction Records: Resources for Job Seekers, Workers and Employers, \url{https://www.eeoc.gov/arrestandconviction}.

\bibitem{chainitpay} {\color{black}\href{https://chainit.com/product/digital-pay/}{ChainIT Pay} official product materials, About ChainIT Pay and Money Movement, describing ChainIT Pay as the ChainIT ecosystem's money-movement layer across supported bank, card, instant-payment, wallet, and stablecoin rails; integration varies by deployment, \url{https://chainitpay.com/about/} and \url{https://chainitpay.com/money-movement/}.}

\bibitem{provable-compliance} ChainIT, \href{https://chainit.com/white-papers/}{Provable Compliance}: A Protocol for Runtime Compliance Evidence and Transaction-Specific Decisioning, CIT-PC-WP-3.0, companion white paper, including modular Evidence \href{https://chainit.com/product/data-tokenization/}{VDTs} and source classes, the Compliance Policy Profile, the Compliance Decision VDT, Workflow Engine disposition configuration, source-specific freshness, and the May Execute predicate, \url{https://chainit.com/white-papers/}; arXiv:\href{https://arxiv.org/abs/2610.22913}{2610.22913}.

\bibitem{vcdm2} W3C Recommendation, Verifiable Credentials Data Model v2.0 (May 15, 2025), \url{https://www.w3.org/TR/vc-data-model-2.0/}.

\bibitem{did} W3C Recommendation, Decentralized Identifiers (DIDs) v1.0 (July 19, 2022), \url{https://www.w3.org/TR/did-core/}.

\bibitem{rfc6749} IETF RFC 6749, The OAuth 2.0 Authorization Framework, \url{https://www.rfc-editor.org/rfc/rfc6749.html}.

\bibitem{rfc9396} IETF RFC 9396, OAuth 2.0 Rich Authorization Requests, \url{https://www.rfc-editor.org/rfc/rfc9396.html}.

\bibitem{rfc8693} IETF RFC 8693, OAuth 2.0 Token Exchange, \url{https://www.rfc-editor.org/rfc/rfc8693.html}.

\bibitem{rfc9635} IETF RFC 9635, Grant Negotiation and Authorization Protocol (GNAP), \url{https://www.rfc-editor.org/rfc/rfc9635.html}.

\bibitem{dennis1966} J. B. Dennis and E. C. Van Horn, Programming Semantics for Multiprogrammed Computations, Communications of the ACM 9(3):143–155, 1966.

\bibitem{rfc2693} IETF RFC 2693, SPKI Certificate Theory, \url{https://www.rfc-editor.org/rfc/rfc2693.html}.

\bibitem{blaze1996} M. Blaze, J. Feigenbaum, and J. Lacy, Decentralized Trust Management, Proc. IEEE Symposium on Security and Privacy, pp.~164–173, 1996.

\bibitem{macaroons} A. Birgisson, J. G. Politz, {\'U}. Erlingsson, A. Taly, M. Vrable, and M. Lentczner, Macaroons: Cookies with Contextual Caveats for Decentralized Authorization in the Cloud, Proc. Network and Distributed System Security Symposium (NDSS), 2014.

\bibitem{clarkwilson} D. D. Clark and D. R. Wilson, A Comparison of Commercial and Military Computer Security Policies, Proc. IEEE Symposium on Security and Privacy, pp.~184–194, 1987.

\bibitem{sandhu1996} R. S. Sandhu, E. J. Coyne, H. L. Feinstein, and C. E. Youman, Role-Based Access Control Models, IEEE Computer 29(2):38–47, 1996.

\bibitem{landrock1998} P. Landrock and T. Pedersen, WYSIWYS? What You See Is What You Sign?, Information Security Technical Report 3(2):55–61, 1998.

\bibitem{psd2rts} Commission Delegated Regulation (EU) 2018/389 supplementing Directive (EU) 2015/2366 with regard to regulatory technical standards for strong customer authentication and common and secure open standards of communication, Article 5 (dynamic linking), OJ L 69, 13.3.2018.

\bibitem{spiffe} Cloud Native Computing Foundation, SPIFFE: Secure Production Identity Framework for Everyone, \url{https://spiffe.io/}.

\bibitem{rfc9334} IETF RFC 9334, Remote ATtestation procedureS (RATS) Architecture, \url{https://www.rfc-editor.org/rfc/rfc9334.html}.

\bibitem{nist207} NIST Special Publication 800-207, Zero Trust Architecture (August 2020).

\bibitem{south2025} T. South, S. Marro, T. Hardjono, R. Mahari, C. D. Whitney, D. Greenwood, A. Chan, and A. Pentland, Authenticated Delegation and Authorized AI Agents, arXiv:2501.09674, 2025.

\bibitem{chan2024ids} A. Chan et al., IDs for AI Systems, arXiv:2406.12137, 2024.

\bibitem{ap2} Google, Agent Payments Protocol (AP2), announced September 2025; specification at \url{https://ap2-protocol.org/ap2/specification/} (accessed October 2026).

\bibitem{acp} OpenAI and Stripe, Agentic Commerce Protocol (ACP), announced September 2025, \url{https://openai.com/index/buy-it-in-chatgpt/}.

\bibitem{visatap} Visa, Trusted Agent Protocol, developed with Cloudflare, announced October 2025, \url{https://investor.visa.com/news/news-details/2025/Visa-Introduces-Trusted-Agent-Protocol-An-Ecosystem-Led-Framework-for-AI-Commerce/}.

\bibitem{lan2026} Q. Lan, A. Kaul, S. Jones, and S. Westrum, Zero-Trust Runtime Verification for Agentic Payment Protocols: Mitigating Replay and Context-Binding Failures in AP2, arXiv:2602.06345, 2026.

\bibitem{crosby2009} S. A. Crosby and D. S. Wallach, Efficient Data Structures for Tamper-Evident Logging, Proc. 18th USENIX Security Symposium, pp.~317–334, 2009.

\bibitem{rfc9162} IETF RFC 9162, Certificate Transparency Version 2.0, \url{https://www.rfc-editor.org/rfc/rfc9162.html}.

\end{thebibliography}
\end{document}